\documentclass[dvipdfmx]{emulateapj-rtx4}

\usepackage{graphicx}

\usepackage{apjfonts}

\usepackage{amssymb}
\usepackage{amsmath}
\usepackage{color}
\usepackage{ulem}
\usepackage{bm}
\usepackage{multirow}

\newcommand{\ltsim}{\protect\raisebox{-0.5ex}{$\:\stackrel{\textstyle <}{\sim}\:$}}
\newcommand{\gtsim}{\protect\raisebox{-0.5ex}{$\:\stackrel{\textstyle >}{\sim}\:$}}

\slugcomment{Received 2015 Jannuary 29; accepted 2015 May 4; published 2015 MM DD}

\shorttitle{Merger of White Dwarfs}
\shortauthors{Sato et al.}

\begin{document}

\title{A systematic study of carbon--oxygen white dwarf mergers:
mass combinations for Type I\lowercase{a} supernovae}

\author{Yushi Sato$^{1,2}$, Naohito Nakasato$^{3}$, Ataru Tanikawa$^{3,4}$,
Ken'ichi Nomoto$^{5,7}$,
Keiichi Maeda$^{5,6}$, Izumi Hachisu$^{2}$}
\affil{$^{1}$Department of Astronomy, Graduate School of Science,
The University of Tokyo, 7-3-1 Hongo, Bunkyo-ku, Tokyo 113-0033, Japan\\
$^{2}$Department of Earth Science and Astronomy, College of Arts and Sciences,
The University of Tokyo, 3-8-1 Komaba, Meguro-ku, Tokyo 153-8902, Japan
; sato@ea.c.u-tokyo.ac.jp\\
$^{3}$Department of Computer Science and Engineering, University of Aizu, 
Tsuruga Ikki-machi Aizu-Wakamatsu, Fukushima 965-8580, Japan\\
$^{4}$RIKEN Advanced Institute for Computational Science, 7-1-26,
Minatojima-minami-machi, Chuo-ku, Kobe, Hyogo 650-0047, Japan\\
$^{5}$Kavli Institute for the Physics and Mathematics of the Universe (WPI),
The University of Tokyo, 5-1-5 Kashiwanoha, Kashiwa, Chiba 277-8583, Japan\\
$^{6}$Department of Astronomy, Kyoto University, Kitashirakawa-Oiwake-cho,
Sakyo-ku, Kyoto 606-8502, Japan}

\altaffiltext{7}{Hamamatsu Professor}

\begin{abstract}
Mergers of two carbon--oxygen (CO) WDs have been
considered as progenitors of Type Ia supernovae (SNe Ia).
Based on smoothed particle hydrodynamics (SPH) simulations,
previous studies claimed that mergers of CO WDs lead to an SN Ia explosion
either in the dynamical merger phase or stationary rotating merger
remnant phase. However, the mass range of CO WDs
that lead to an SN Ia has not been clearly identified yet.
In the present work, we perform systematic SPH merger simulations for
the WD masses ranging from $0.5~M_{\odot}$ to 
$1.1~M_{\odot}$ with higher resolutions than the previous
systematic surveys and examine whether or not carbon burning occurs
dynamically or quiescently in each phase.
We further study the possibility of SN Ia explosion
and estimate the mass range of CO WDs that lead to an SN Ia.
We found that when the both WDs are massive, i.e., 
in the mass range of $0.9~M_{\odot} {\le} M_{1,2} {\le} 1.1~M_{\odot}$, 
they can explode as an SN Ia in the merger phase. 
On the other hand, when the more massive WD is in the range of
$0.7~M_{\odot} {\le} M_{1} {\le} 0.9~M_{\odot}$ and 
the total mass exceeds $1.38~M_{\odot}$, 
they can finally explode in the stationary rotating merger remnant phase.
We estimate the contribution of CO WD mergers to the entire SN Ia rate
in our galaxy to be of ${\ltsim} 9\%$.
So, it might be difficult to explain all galactic SNe Ia by CO WD mergers.
\end{abstract}



\section{Introduction}
\label{introduction}

SNe Ia play the important roles in the determination
of cosmological parameters as luminous standard candles 
\citep[e.g.,][]{riess, perlmutter} and in the chemical evolution of galaxies 
as major sources of iron group elements \citep[e.g.,][]{kobayashi}.
However, their progenitors are still not identified yet \citep[e.g.,][]{maoz}.
They are considered as a thermonuclear explosion of a CO WDs 
in a binary system, in which the WD accretes mass from
its companion and the WD mass approaches the Chandrasekhar mass 
($\mathrm{M_{Ch}} \sim 1.4~M_{\odot}$). 
But, it is still controversial whether its companion is a non-degenerate
star, i.e., single degenerate (SD) model, or a degenerate star, i.e., 
double degenerate (DD) model. In the SD model, a CO WD accretes 
hydrogen/helium-rich gas from the companion and increases its mass upto 
$M_\mathrm{Ch}$.  Finally, carbon burning starts at the center of 
the CO WD and explodes as an SN Ia \citep{whelan, nomoto82, hkn96,
hkn99, hknu99}.
On the other hand, in the DD model, both the components are CO WDs.
Because binaries lose their orbital angular momentum by emitting
gravitational waves, they will eventually merge. If their total mass exceeds 
$M_\mathrm{Ch}$, the binary finally explodes as an SN Ia 
\citep{iben,webbink}. 

Some clues on the progenitors of SNe Ia were found in recent
observations.  In particular, neither surviving companions nor signatures
of them were detected in some SNe Ia \citep[e.g.,][]{schaefer12}.
This fact supports the DD system, 
although no detection of companions can also be explained
by the SD model \citep{distefano, justhum, hkn12}.
In some SNe Ia, signatures of circumstellar matter (CSM) 
were detected \citep[e.g.,][]{patat07, blondin09, simon09, sternberg11}.
This supports the SD model \citep[e.g.,][for a recent
review]{maoz}, although this can be explained by 
some DD models \citep{raskin13, shen13, soker}. 
Observations do still not clarify which model is the main
progenitors of SNe Ia, i.e., SD or DD (or other type) system.

The DD model has a theoretical difficulty as a progenitor model.
Some theoretical studies indicated that CO WD mergers can not become
an SN Ia, but collapse to a neutron star \citep[e.g.,][]{saio, saio04,
kondo}. These studies calculated only the evolution of merger remnants 
after DD systems merged. Their calculations were one-dimensional 
(1D) spherically symmetric and assumed stationary state.  
However, the merger of a DD system is a three-dimensional (3D)
and dynamical event, so multi-dimensional hydrodynamical simulations
are necessary to reach a definite conclusion. 
\citet{benz} used 3D smoothed particle hydrodynamics (SPH) 
code and simulated mergers of two CO WDs.
Following this work,
there were several studies on mergers of DD systems
\citep{rasio, segretain, guerrero, yoon, loren, fryer, pakmor10, 
dan11, raskin12, raskin14, moll}.
These studies concluded that some DD pairs can explode as SNe Ia.
Such successful models can be divided by the dynamical phase when the SN Ia
explosion occurs.

\citet{pakmor10} simulated mergers of massive CO WDs (${\sim}0.9~
{M_{\odot}}$) and found that carbon detonation initiates during the
dynamical merger phase because of the compressional heating
by the disrupted secondary
which violently accretes onto the primary.  The binary system finally
explodes as a subluminous SN Ia.  \citet{pakmor12a} also simulated 
mergers of more massive CO WDs ($1.1+0.9~{M_{\odot}}$)
and found that the system leads to a normal SN Ia. 

If carbon detonation does not initiate in the dynamical merger phase,
the remnant undergoes three phases.  The first phase is 
the early remnant phase \citep[e.g.,][]{shen12, kashyap}, 
$100$--$1000$~s after the secondary is completely disrupted.
In this phase, the merger remnant does not reach a quasi-stationary state 
yet and still has small non-axisymmetric structures.
The second phase is the viscous evolution phase \citep{schwab, shen12, ji}, 
$10^4$--$10^{8}$~s after merging.
In this phase, the remnant reaches a quasi-stationary, axisymmetric state and 
it evolves in a viscous timescale.
The third is the thermal evolution phase \citep{saio, saio04, yoon, shen12},
${\gtsim}10^3$~years. 
If off-center carbon burning occurs in these three phases
before the rotating core of the remnant reaches $M_{\rm Ch}$,
it likely converts the CO WD to an oxygen--neon--magnesium (ONeMg) WD.
The ONeMg WD finally collapses to a neutron star when its core mass
reaches $M_{\rm Ch}$.  \citet{yoon} simulated the merger of CO WDs with masses
of $0.9+0.6~M_{\odot}$ and followed the evolution of the merger remnant.
They found that the remnant can avoid off-center carbon 
burning and explodes as an SN Ia in the thermal evolution phase,
if it satisfies several conditions.

Thus, these works concluded that some DD systems can become SNe Ia.
However, the mass range of CO WDs that lead to SNe Ia is not clarified yet.
In this work, we simulate mergers of CO WDs with a mass range of
$0.5$--$1.1~{M_{\odot}}$ with our SPH code until the end of the early remnant phase,
and identify the mass range of CO WDs that lead to SNe Ia.
Using the obtained mass range of CO WDs, we estimate
the contribution of CO WD mergers to the entire SNe Ia in our galaxy.

Similar parameter surveys have already been done by 
\citet{dan12,dan14} and \citet{zhu}. \citet{zhu} mainly focused
on the (early) remnant phase and the resolutions of their simulations 
are lower than our study.
Although \citet{dan12,dan14} covered the merger phase and 
remnant phase, their numerical resolution is much lower than ours.
We expect that SN Ia explosions occur not only in the early remnant phase
but also in the dynamical merger phase.  In the present work,
we adopt higher resolution than two times those of the previous works 
\citep[e.g.,][]{zhu}.  This is because the numerical resolution is
one of the most important parameters to identify the initiation of detonation
in the dynamical merger phase \citep{pakmor12b}.  Therefore,
we adopt four different resolutions and check the numerical convergence. 

To investigate the possibility that carbon burning leads to an SN Ia, 
we check the density and temperature of SPH particles
and identify carbon burning by the condition of
$\tau_{\rm CC} < \tau_{\rm cool}$, where $\tau_{\rm CC}$ is the 
timescale of carbon burning and $\tau_{\rm cool}$ is the cooling timescale.
In the dynamical merger phase, $\tau_{\rm cool}$ is the dynamical 
timescale of adiabatic expansion.
On the other hand, in the (early) remnant phase, $\tau_{\rm cool}$ is 
the timescale of neutrino cooling.
These are necessary but not sufficient conditions
for a thermonuclear explosion.
In this sense, our results would not be conclusive
but are sufficiently suggestive.
Since the temperature has numerical noise in our SPH simulation,
we have to treat the temperature carefully.
This noise comes from fluctuations of density and internal energy,
which arise due to numerical resolution, finite neighbor particles,
and the accuracy of time-integration.
We use both the raw and smoothed temperatures to estimate the noise
in our SPH simulation \citep{dan12}.
The raw temperature is the temperature
that our SPH code originally generates.
The smoothed temperature is an averaged temperature 
calculated from neighbor particles, 
which could avoid large numerical noises
(defined in Section \ref{mereger_phase}).
In the present version of our SPH simulation,
we do not include nuclear reactions
because they are so sensitive to temperature noises
and possibly enhance them erroneously.

This paper is organized as follows. Section \ref{method}
briefly describes our numerical methods.  In section \ref{result},
we show our results and then estimate the rate of SNe Ia
coming from the DD merger systems
and their contribution to the entire SNe Ia in our galaxy. 
In section \ref{discussion}, we compare our results with previous works and 
discuss the dependence of our numerical results on the resolution. 
Finally we conclude the present work in Section \ref{conclusions}.


\section[]{Methods}
\label{method}
Here, we present a brief summary of our numerical method. The details
of the numerical method are already described in \citet{nakasato} and \citet{tanikawa}.

\subsection{Numerical Code}
\label{numerical_method}

SPH is the Lagrangian mesh-free particle
method developed for the study of astrophysical fluid phenomena \citep{lucy, gingold}.
Nowadays, it is applied to more various subjects, e.g., engineering,
meteorology, and Hollywood movies. Recent good reviews about SPH codes are
available, e.g., in \citet{monaghan05} and \citet{rosswog09}.
Our formulation is the one called "vanilla ice" SPH formulation
in \citet{rosswog09}.  Our basic SPH equations consist of
the equation of continuity, equation of motion,
and energy equation for self-gravitating fluid in Lagrangian formulation.

We use "OcTree On OpenCL"(OTOO) code for our 3D simulations
of CO WD mergers, which is developed for various particle simulations of
astrophysical fluid phenomena \citep{nakasato}. This code implements
the octree method \citep{barnes} to calculate gravity and neighbor
particles for SPH. It is optimized for multiple CPUs and GPUs
in heterogeneous computational resources.

In this code, we adopt the 3rd-order spline kernel and its derivative is
modified in the same way as that proposed by \citet{thomas} to avoid the
pairing instability.  The smoothing length is determined to keep the average
number of neighbors being about 75 in every step \citep{thacker}. 
The formulation of the artificial viscosity is basically the same as \citet{monaghan92},
but the viscosity coefficient is time-dependent \citep{morris}.
More detail explanation can be seen in \citet{rosswog00} and 
\citet{monaghan05}.
We also introduce the Balsara switch to shut off the artificial viscosity
in no shock regions \citep{balsara}.
We adopt a leap-frog scheme for time integration.

We use the Helmholtz equation of state (EOS) \citep{timmes}
and assume a uniform chemical composition of 50\%
carbon and 50\% oxygen. This chemical composition is not changed throughout
the merger simulation because we do not include nuclear reactions.

\subsection{Initial Setup}
\label{initial_setup}

The initial setup of our simulations is similar to those of
\citet{rasio} and \citet{dan11}. We separately generate each CO WD 
from spherically symmetric density profiles of a perfectly degenerate star,
and set the temperature at $10^{6}$~K everywhere.
To reduce numerical noise in the mapping from the 1D spherically
symmetric density profile to our 3D SPH density distribution,
we relax the SPH particles for 20 physical seconds with velocity
damping force but without the evolution of internal energy. 
For the damping force, we adopt
\begin{equation}
\frac{d\mathbf{v}_{i}}{dt} = -\frac{\mathbf{v}_{i}}{C_{\rm damp}dt},
\end{equation}
where $\mathbf{v}_{i}$ is the velocity of $i$th particle, $C_{\rm damp}$
is the inverse of relaxation timescale and we fix $C_{\rm damp}=128.0$.

Next, we put two CO WDs to the distance enough
to avoid the Roche lobe over flow (RLOF). 
We assume that they are on a circular
orbit and their spins synchronize each other with the orbital motion.
To finish setting the initial condition, we gradually
decrease the separation and relax them with the above damping force 
and the evolution of internal energy in a co-rotating frame.
When an SPH ($i$th) particle of the secondary approaches
closely enough the L1 point, i.e., 
\begin{equation}
\|\mathbf{r}_{{\rm sec},i} - \mathbf{r}_{\rm L1}\| < 0.2R_{2},
\end{equation}
we stop the calculation.
Here $\mathbf{r}_{{\rm sec},i}$ is the position of $i$th particle of the
secondary, $\mathbf{r}_{\rm L1}$ is that of L1 point, and
$R_{2}$ is the effective radius of the secondary.

Finally, we transfer the SPH particles from the co-rotating frame 
to a rest frame.


\section{Results}
\label{result}

We summarize the results of our simulations in Table A1.
Because we aim to identify the mass range of CO WDs that lead
to an SN Ia, our initial models cover the entire mass range of CO WDs,
i.e., from $M_{\rm WD}=0.5$ to $1.1~M_{\odot}$.  We prepare binary
models of 0.5, 0.6, 0.7, 0.8, 0.9, 1.0, and $1.1~M_{\odot}$, and perform
the simulations of mergers for all 28 mass combinations. 
Since we also investigate the dependence of our results on the numerical
resolution, we perform the same simulations with different resolutions, 
which are $10k, ~50k, ~100k, ~500k$ SPH particles per one solar mass
(here $k~\equiv~1,024$). 
All simulations were performed from the start of RLOF until
the formation of almost stationary remnant, several hundred seconds 
after the secondary is completely disrupted. 
We divide the dynamical evolution of a merger into the following two phases.
(1) The merger phase is the period during which the secondary 
dynamically accretes onto the primary.
(2) The (early) remnant phase is from the complete disruption of 
the secondary to the stage that the system reaches almost 
a quasi-stationary state.

\begin{figure*}
 \begin{center}
  \includegraphics[width=16cm, height=8cm, clip]{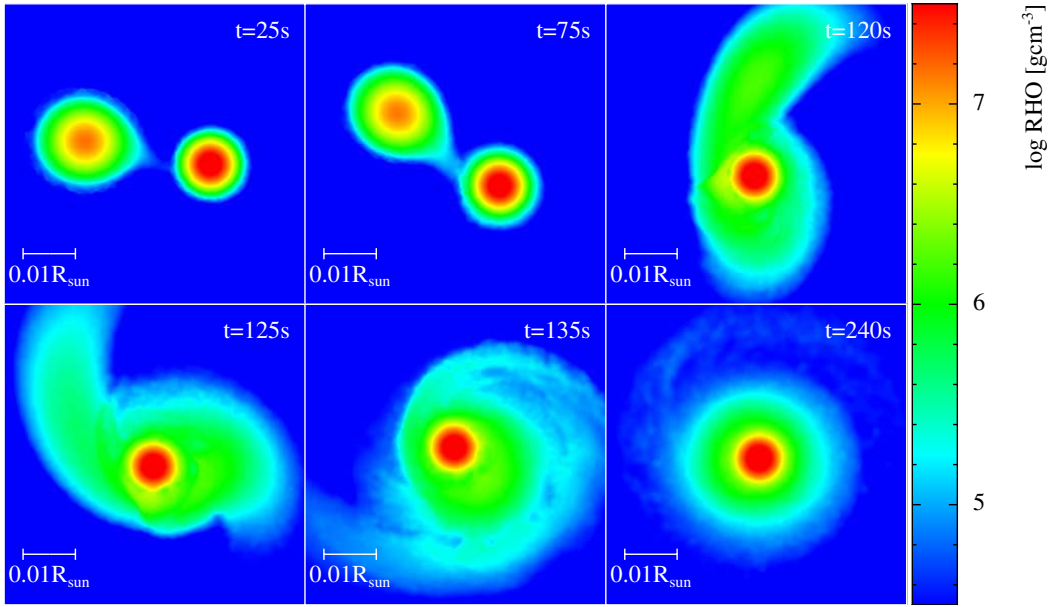}
 \end{center}
 \vspace{10pt}
 \caption{Density profiles in the equatorial plane for the dynamical
evolution of our merger simulation.  The mass combination is
$1.1+0.9~M_{\odot}$, and the resolution is $500k~M_{\odot}^{-1}$.
Colors indicate density in a logarithmic scale.}
 \label{fig.3-1}
\end{figure*}

Figure \ref{fig.3-1} shows the density profiles in the equatorial 
plane for an example of our simulations, whose mass combination is 
$1.1+0.9~M_{\odot}$ and resolution is $500k~M_{\odot}^{-1}$.
The merger phase covers the first ($t~=~25$ s)
to fifth ($t~=~135$ s) panels, and the early remnant phase corresponds to
the sixth ($t~=~240$ s) panel.
Their morphological structures are consistent with the previous works 
\citep[e.g.,][]{pakmor12a}.

First we check whether dynamical carbon burning starts or not
in the merger phase for all simulations because it is a necessary
condition for an SN Ia explosion.  Then, we check steady carbon
burning in the early remnant phase.  If it occurs in the early remnant
phase, carbon burning converts
a CO WD into an ONeMg WD and the merger remnant will not become an SN Ia. 
Although we perform simulations with four different resolutions,
we first focus on the highest resolutions ($= 500k ~M_{\odot}^{-1}$).

\subsection{Merger Phase}
\label{mereger_phase}

\citet{pakmor10} first suggested that CO WD mergers lead to an SN Ia
in the merger phase. They called such a model the (carbon-ignited)
violent merger (VM) scenario.
In this scenario, matter of the secondary violently accretes
onto the primary and such violent accretion causes dynamical carbon burning.
As a result, detonation wave would be formed and propagate into
the primary with converting its carbon--oxygen into iron group elements.
Finally, the system explodes as an SN Ia.  Because our simulations cannot
directly resolve the initiation of detonation, we try to judge the
occurrence of dynamical carbon burning in the merger phase.
This condition is a necessary condition for an SN Ia explosion.

For this purpose, we extract the highest temperature particle
in the merger phase for all simulations.
The condition for dynamical carbon burning is
\begin{equation}
\tau_{\rm CC} < \tau_{\rm dyn},
\label{dynamical_burning}
\end{equation}
where $\tau_{\mathrm{CC}}$ is a carbon burning timescale defined by
\begin{equation}
\tau_{\rm CC} = {{C_P T} \over \epsilon_{\rm CC}},
\end{equation}
and $\tau_{\mathrm{dyn}}$ is a dynamical timescale \citep{nomoto82}
defined by
\begin{equation}
\tau_{\rm dyn} = {1 \over \sqrt{24{\pi}G\rho}},
\end{equation}
where $C_P$ is the specific heat
at constant pressure, $\epsilon_{\rm CC}$ is the energy generation
rate of carbon burning.  We calculate the both timescales
for each particle in the merger phase and examine whether the particles
satisfy Equation (\ref{dynamical_burning}).
In this work, $C_P$ is derived from the Helmholtz EOS of \citet{timmes}
and $\epsilon_{\rm CC}$ is the same as that of \citet{dan14},
originally proposed by \citet{blinnikov}. Its formulation is
\begin{equation}
\epsilon_{\rm CC} = {\rho}~q_{\rm C} ~A_{\rm T}
~Y_{\rm C}^{2}\exp (-Q/T_{\rm 9a}^{1/3}+f_{\rm CC}),
\end{equation}
where $q_{\rm C} = 4.48{\times}10^{18}${\,}erg{\,}mol$^{-1}$
\citep{blinnikov}, $A_{\rm T} = 8.54{\times}10^{26} T_{\rm 9a}^{5/6}
T_{9}^{-3/2}${\,}s$^{-1}${\,}mol$^{-1}${\,}cm$^3$, $T_{9} \equiv T/10^{9}${\,}K,
$T_{\rm 9a} \equiv T_{9}/(1 + 0.067 T_{9})$, $Q = 84.165$ \citep{fowler}.
The carbon abundance is calculated as 
$~Y_{\rm C} = n_{\rm C}/({\rho}N_{\rm a}) = 0.033${\,}mol{\,}g$^{-1}$, 
where $n_{\rm C}$ is the number density of carbon and 
$N_{\rm a}$ is the Avogadro number. 
A screening factor $f_{\rm CC}$ is ignored here,
because we focus on the start of dynamical carbon burning and
the factor of self-acceleration for nuclear burning described
in \citet{frank} is not applied to the initiation of carbon burning.

\begin{figure}
  \begin{center}
    \begin{minipage}{0.95\hsize}
      \begin{center}
       \includegraphics[clip, width=7.5cm, angle=0]{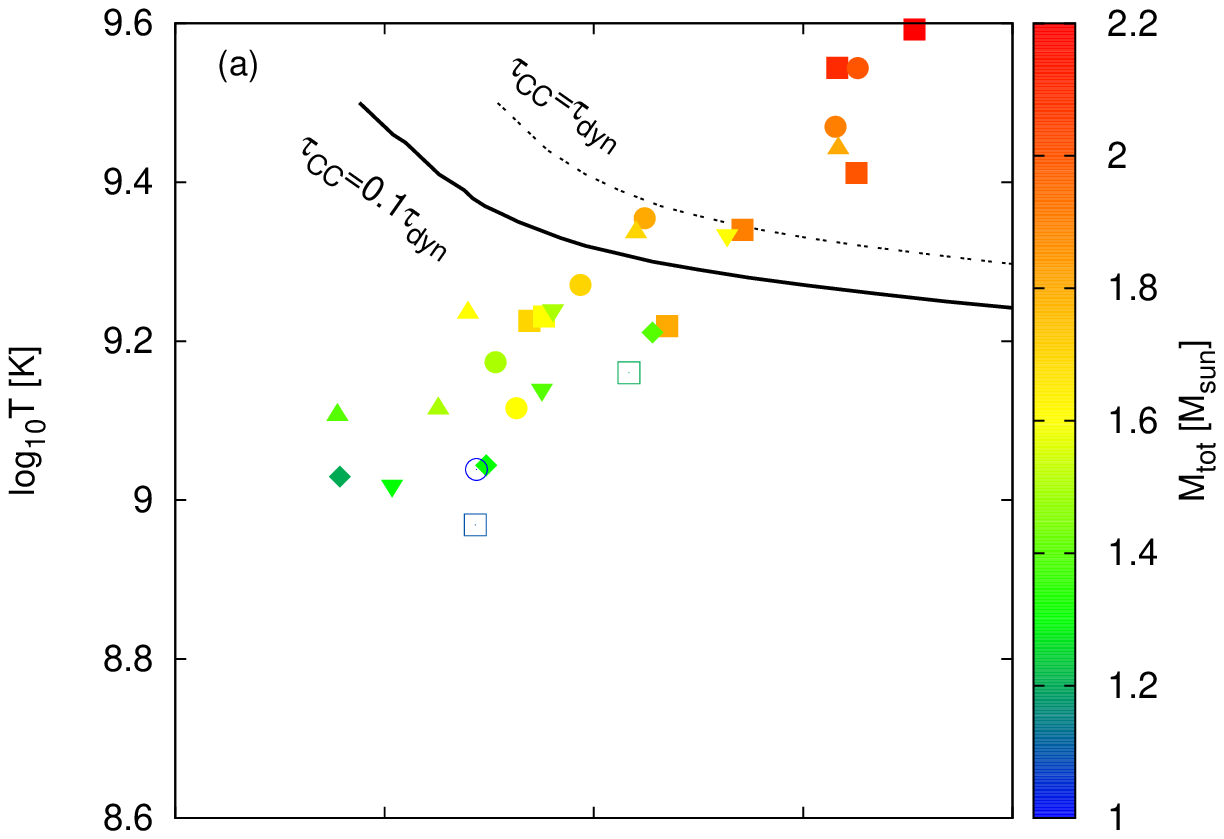}
      \end{center}
    \end{minipage}

    \begin{minipage}{0.95\hsize}
      \begin{center}
        \includegraphics[clip, width=7.5cm, angle=0]{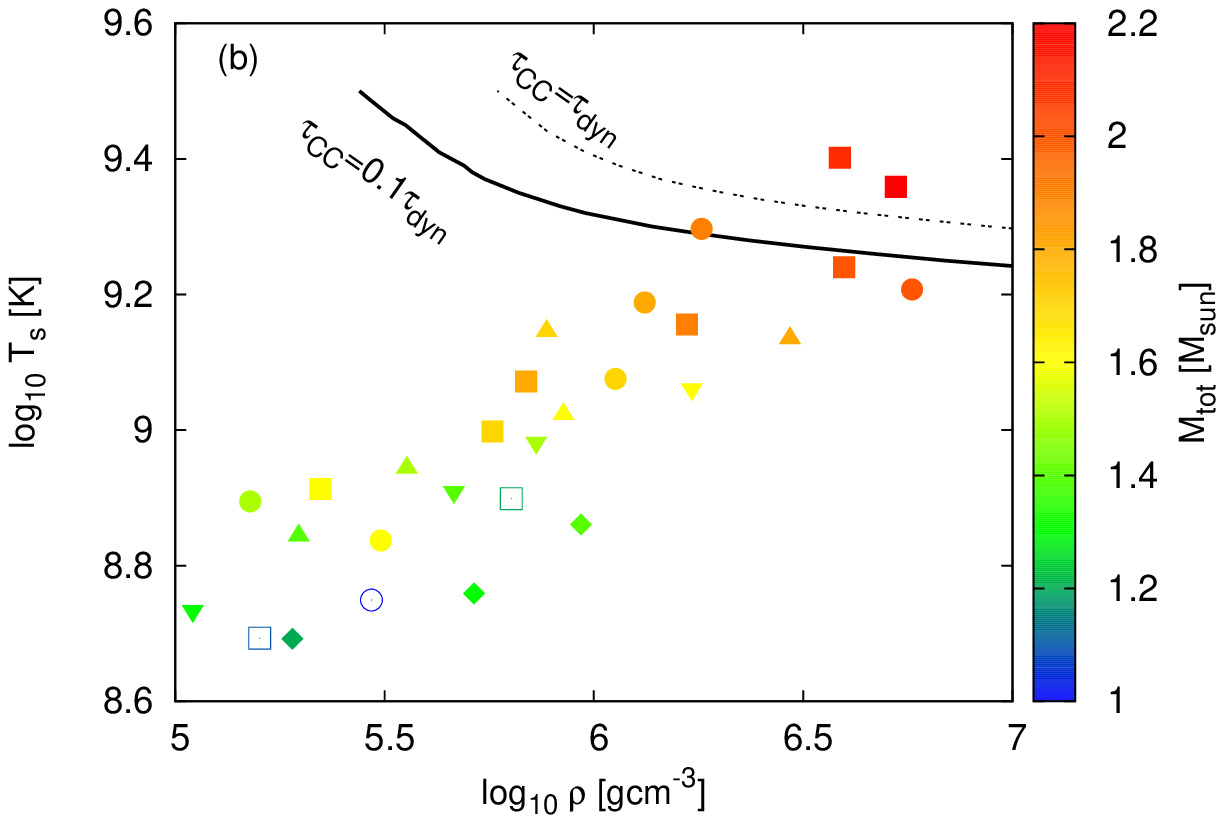}
      \end{center}
    \end{minipage}

  \end{center}
  \vspace{10pt}
 \caption{Density and temperature of the highest temperature particle
in the merger phase for all mass combinations. Numerical resolutions
of these models are $500k~M_{\odot}^{-1}$. Colors of symbols indicate
the total mass of the system as indicated in the rightside of the figures,
and shapes of symbols do the mass of the primary. 
Filled squares are the $1.1~M_{\odot}$ primary, filled circles 
$1.0~M_{\odot}$, filled triangles $0.9 ~M_{\odot}$, filled inverted
triangles $0.8~M_{\odot}$, filled diamonds $0.7~M_{\odot}$, 
open squares $0.6~M_{\odot}$, open circles $0.5~M_{\odot}$.
Solid lines indicate $\tau_{\rm CC} = \tau_{\rm dyn}$,
and dashed lines do $\tau_{\rm CC} = 0.1\tau_{\rm dyn}$.  
(a) Raw temperature of SPH particles. (b) Smoothed temperature, $T_{\rm s}$, 
of SPH particles defined by Equation (\ref{smoothed_temperature}).}
 \label{fig.3-2}
\end{figure}

Figure \ref{fig.3-2} shows the density and temperature of
a particle with the highest temperature in the merger phase for all mass
combinations (their resolutions are $500k~M_{\odot}^{-1}$),
similar to Figure 12 of \citet{dan14}. The shapes and colors of symbols
indicate the primary's mass and the total mass, respectively.  
Solid lines indicate $\tau_{\rm CC} = \tau_{\rm dyn}$,
and dashed lines do $\tau_{\rm CC} = 0.1\tau_{\rm dyn}$.
For the mass combinations above the solid line, 
dynamical carbon burning occurs in the merger phase.
So, the merger of CO WDs would lead to an SN Ia explosion. 

In our SPH simulation, physical raw temperature of each particle
has numerical noise. 
So we adopt another definition of temperature to reduce the effect of noise,
i.e., smoothed temperature. It is defined by
\begin{equation}
T_{s,i} = \sum_j\frac{m_j}{\rho_j}T_{j}W(r_{ij},h_{ij}),
\label{smoothed_temperature}
\end{equation}
where $m_j$, $\rho_j$, $T_j$ are the mass, density, temperature of 
$j$th particle, respectively. $r_{ij} = \|\mathbf{r}_{i}-\mathbf{r}_{j}\|$,
and $h_{ij} = (h_{i}+h_{j})/2$ is the average of the smoothing lengths
of $i$th and $j$th particles. 

Figure \ref{fig.3-2}(a) and \ref{fig.3-2}(b) show the results for 
raw temperature and smoothed temperature thus defined, respectively.
It is clear that all symbols in Figure \ref{fig.3-2}(a)
move to a lower place in Figure \ref{fig.3-2}(b).
As a result, the number of mass combinations above the solid line 
of $\tau_{\rm CC} = \tau_{\rm dyn}$ decreases.
Figure \ref{fig.3-2}(b) of smoothed temperature shows
that the mass combinations above the solid line would certainly
trigger dynamical carbon burning in the merger phase.

\subsection{Remnant Phase}
\label{remnant_phase}

\begin{figure}
  \begin{center}
    
      \begin{minipage}{0.95\hsize}
      \begin{center}
        \includegraphics[width=9.5cm, height=3.5cm, clip]{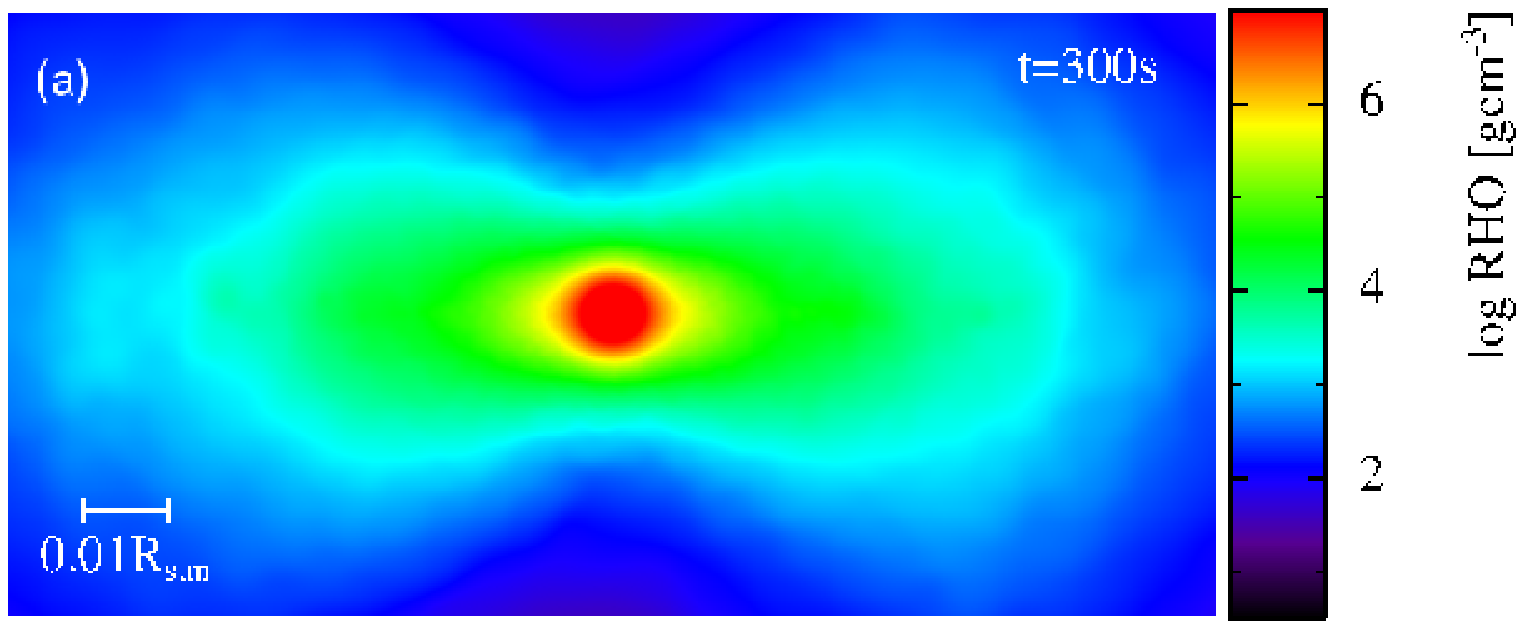}
      \end{center}
      \end{minipage}

      \begin{minipage}{0.95\hsize}
      \begin{center}
        \includegraphics[width=9.5cm, height=3.5cm, clip]{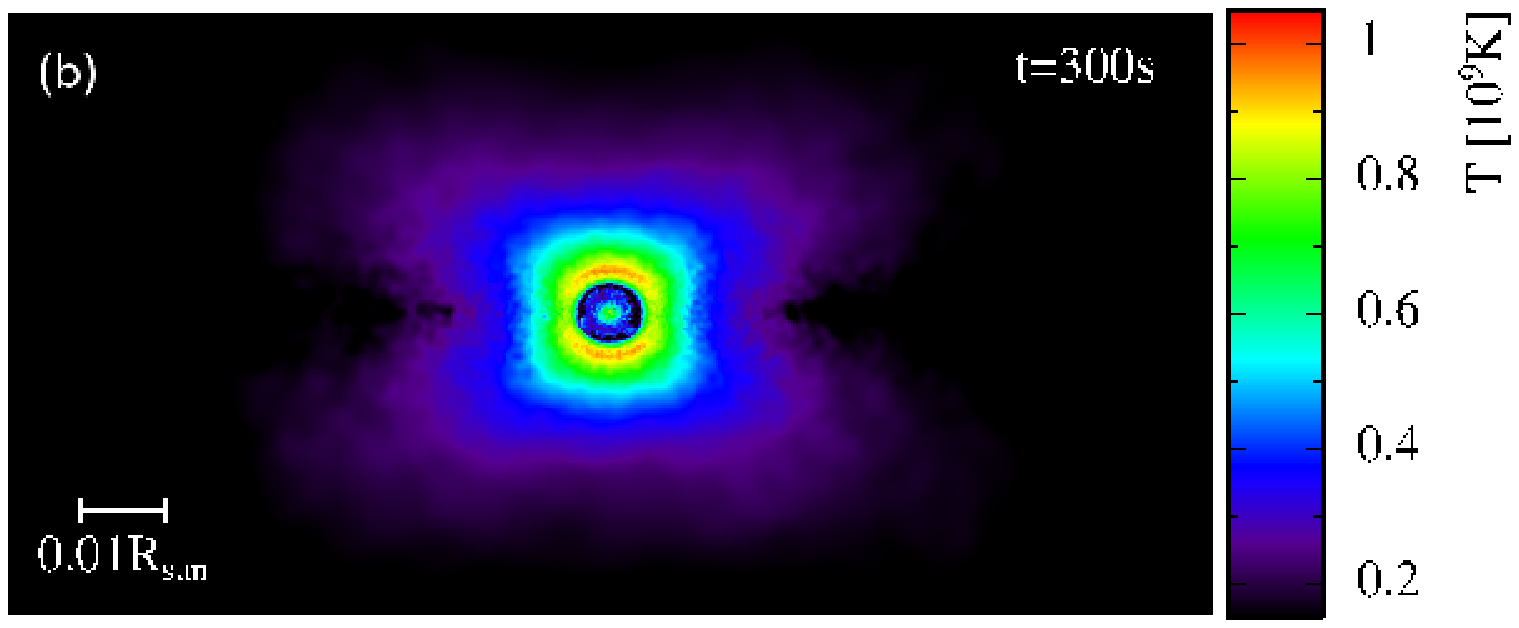}
      \end{center}
      \end{minipage}

  \end{center}
  \vspace{10pt}
 \caption{Structure of a merger remnant at about 
$200$~s (${\sim}7.6~P_{\rm orb, init}$)
after the secondary is disrupted completely.
Here, $P_{\rm orb, init}$ is the initial orbital period.
Its mass combination is 
$1.1+0.9 ~M_{\odot}$ and the numerical resolution is $500k~M_{\odot}^{-1}$,
which is the same model as in Figure \ref{fig.3-1}.  (a) Density profile
(in a logarithmic scale of g~cm$^{-3}$) in the $x--z$ plane.
(b) Temperature structure (in a linear scale of $10^9$~K) 
in the $x--z$ plane.  The central part becomes
slightly hot.  This is caused by numerical noise and artificial viscosity,
so it has no physical meaning.}
 \label{fig.3-3}
\end{figure}

If dynamical carbon burning does not occur in the merger phase, 
the merged object goes into a stationary phase, i.e., the (early) remnant phase.
Figure \ref{fig.3-3}(a) and \ref{fig.3-3}(b) 
show the density and temperature, respectively, of the merger remnant
whose mass combination and resolution are $1.1+0.9 ~M_{\odot}$ and 
$500k~M_{\odot}^{-1}$.
The remnant consists of three components, i.e., a cold core,
hot envelope, and outer disk. The structure of such a remnant has been
studied in the several works as already mentioned in Section 
\ref{introduction}.  
Although there are small differences among these previous works, their results
are almost consistent with each other.
Our results are also consistent with these previous results.

It has long been discussed that such a hot envelope gradually accrete
onto a cold core because of angular momentum loss by some mechanisms
(e.g. viscosity or magnetic field).  If off-center carbon burning occurs
quiescently during such an accretion phase, carbon deflagration waves
propagate into the core.  At last, the whole core is converted into 
an ONeMg WD \citep[e.g.,][]{saio, saio98, saio04}.
In such a case, it can not explode as an SN Ia,
even if its total mass exceeds $M_{\rm Ch}$. Instead, it collapses
to a neutron star \citep{kondo}.
On the other hand, if off-center carbon burning
does not occur, the core remains unchanged as a CO WD and
surrounding matter continues
to accrete onto the core.  When the core mass exceeds $M_{\rm Ch}$,
it explodes as an SN Ia.  It is critically important to examine whether
carbon burning starts quiescently off-center in the remnant phase. 

We examine, in the same way as that of dynamical carbon
burning in the merger phase, whether or not off-center carbon burning
occurs in the remnant phase.
Carbon burning quiescently occurs near the boundary between
the cold core and hot envelope, if the condition for carbon burning, i.e.,
\begin{equation}
\tau_{\rm CC} < \tau_\nu,
\label{carbon_burning_neutrino}
\end{equation}
is satisfied in the remnant phase. Here, $\tau_\nu$ is a timescale of
neutrino cooling and we use the description of \cite{itoh} 
for calculating it from the density and temperature of SPH particles.
We find the highest temperature particle in the remnant phase
for all mass combinations, and examine Equation 
(\ref{carbon_burning_neutrino}).
If the highest temperature particle satisfies the condition,
we regard that off-center carbon burning starts and converts the CO core
into an ONeMg core.  Then, the system finally collapses to a neutron star
if the total mass exceeds $M_{\rm Ch}$.
On the other hand, if there are
no particles that satisfy the condition 
and the total mass of the remnant
exceeds $M_{\rm Ch}$, we consider that the remnant becomes an SN Ia. 

\begin{figure}
  \begin{center}

    \begin{minipage}{0.95\hsize}
      \begin{center}
        \includegraphics[width=7.5cm, angle=0]{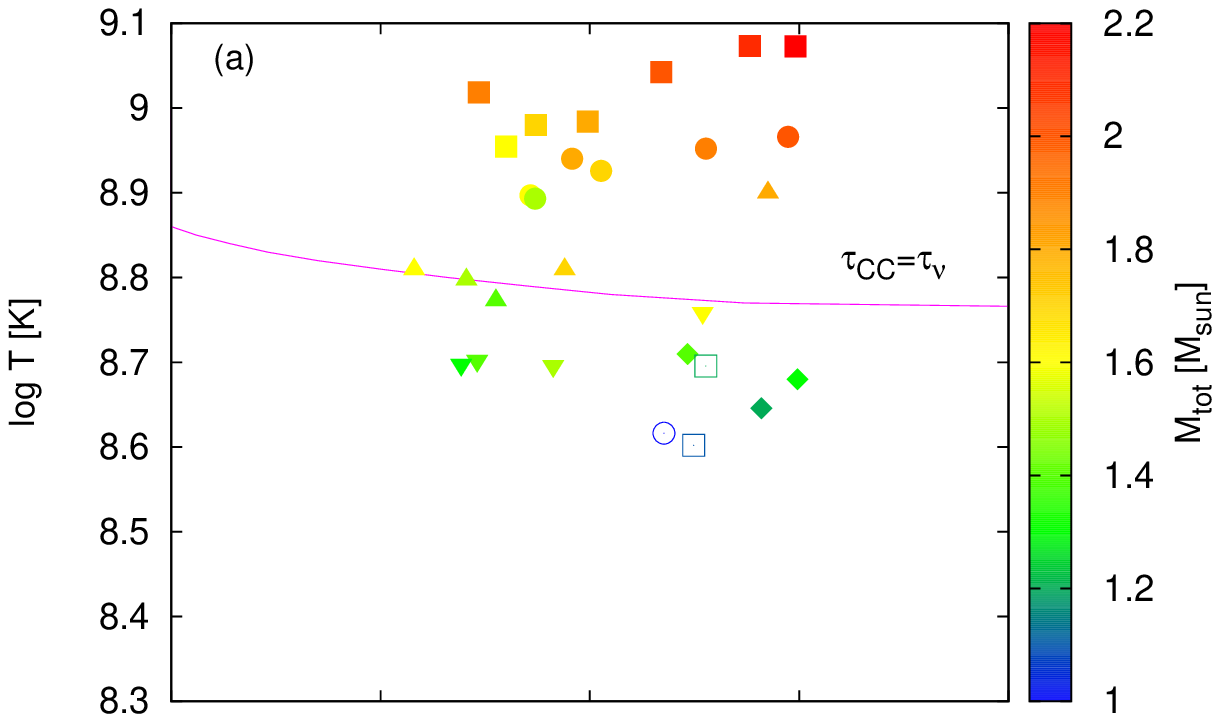}
      \end{center}
    \end{minipage}

    \begin{minipage}{0.95\hsize}
      \begin{center}
        \includegraphics[width=7.5cm, angle=0]{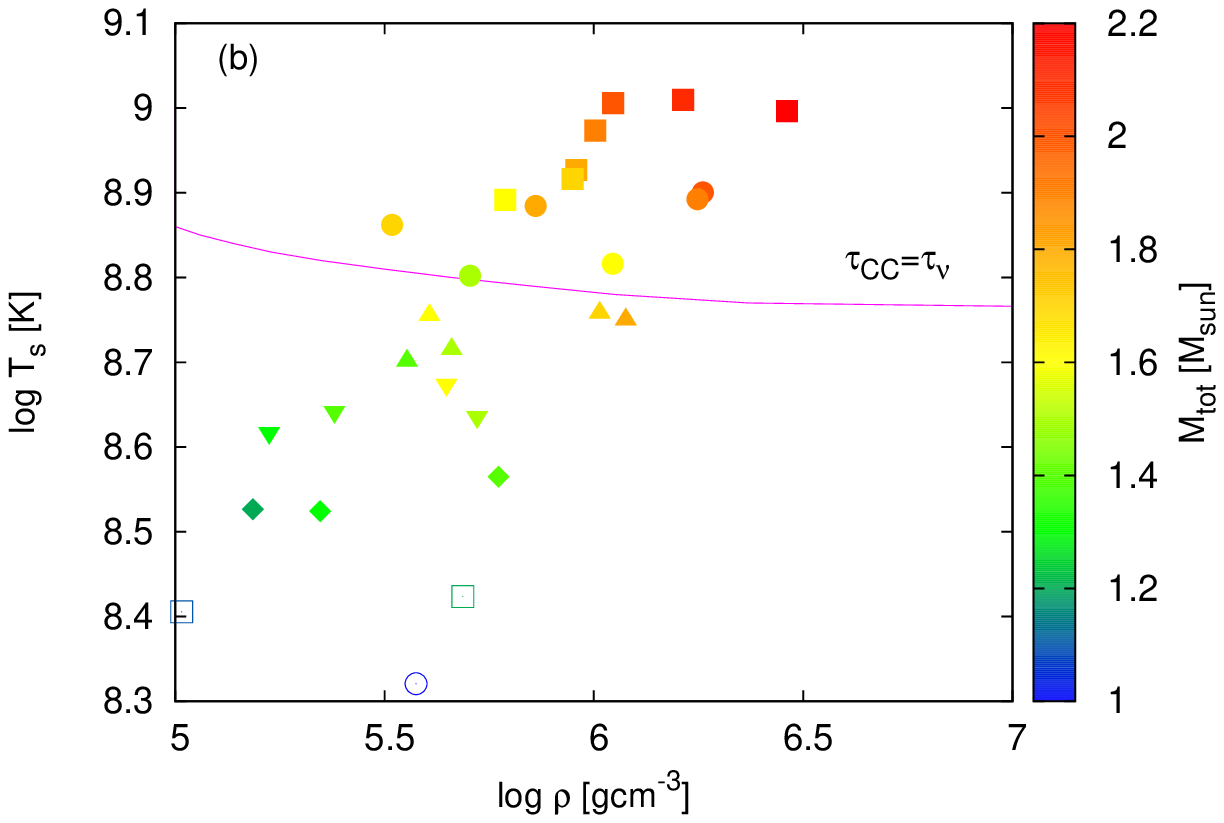}
      \end{center}
    \end{minipage}

  \end{center}
  \vspace{10pt}
 \caption{Same as Figure \ref{fig.3-2}, but for the remnant phase.
Magenta solid lines indicate $\tau_{\rm CC} = \tau_\nu$.}
 \label{fig.3-4}
\end{figure}

Figure \ref{fig.3-4} is the same plot as Figure \ref{fig.3-2}, but for
the remnant phase.  Magenta solid lines indicate $\tau_{\rm CC} = \tau_\nu$.
Figure \ref{fig.3-4}(a) and \ref{fig.3-4}(b) show results of
the raw and smoothed temperatures, respectively.
Off-center carbon burning occurs in the remnant phase for the mass
combination models above the magenta solid line.  They would finally collapse
to a neutron star. On the other hand, the models below the line
would become an SN Ia if the total mass exceeds the Chandrasekhar mass,
i.e., $M_1+M_2 > M_{\rm Ch}$.

We have to follow the viscous and thermal evolution 
phases for obtaining definite conclusion on the off-center carbon burning.
However, we stop our SPH calculation at the end of the early remnant phase
because our SPH code includes no physical viscosities
(e.g., magnetic viscosity).  
In the viscous and thermal evolution phases, 
the hot envelope further accretes onto the cold core
and compress itself on the cold core. The temperature
and density might further increase. 
As a result, carbon ignites off-center even for the cases 
of no off-center burning in the early remnant phase.
In fact, \citet{shen12} and \citet{schwab} followed the evolution of merger remnants
and showed that off-center carbon ignition starts in the viscous and 
thermal evolution phases for some cases.
\citet{yoon} performed SPH simulation of CO WD merger whose mass
combination is $0.9+0.6 ~M_{\odot}$ and further followed the evolution of
the merger remnant with a 1D stellar evolution code.
They found that off-center carbon burning can be avoided
when the highest temperature is lower than the threshold for carbon ignition,
angular momentum loss occurs with a timescale longer than that of
neutrino cooling, and the mass accretion rate is 
$\dot{M}~\leq~5{\times}10^{-6}$ to $10^{-5}~M_{\odot}$~yr$^{-1}$.
Our present condition for off-center carbon burning is posed only
for the (early) remnant phase but not applied yet for 
the viscous and thermal accretion phases. 
In this sense, our results only for the early remnant phase are
not definite answers.  We leave such viscous and thermal evolutions
of the merger remnants in our future works.

\subsection{White Dwarf Mass Combinations For SNe Ia}
\label{mass_conbination}

Now, we have obtained the mass range of CO WDs which
possibly lead to an SN Ia.
Using this mass range of CO WDs, we estimate their contribution
to the entire SNe Ia in our galaxy.  
We consider four paths that CO WD mergers would follow.
The first one is the VM path, the condition of which
is that dynamical carbon burning occurs in the merger phase.
The systems satisfying this condition would explode as an SN Ia
immediately after merging \citep{pakmor10, pakmor11, pakmor12a}.
When dynamical carbon burning does not occur in the merger phase,
the system enters the remnant phase and the disrupted secondary surrounds
the primary. If its total mass exceeds $M_{\rm Ch}$ and off-center
carbon burning does not occur during the accretion phase, carbon burning
occurs at the center of the CO WD and it would finally explode as an SN Ia.
We regard this evolutionary path as the accretion induced explosion (AIE)
path. On the other hand, if off-center carbon burning occurs,
the core of the remnant will be converted into an ONeMg WD and then
collapse to a neutron star when the core mass exceeds $M_{\rm Ch}$.
This is the accretion induced collapse (AIC) path 
\citep{saio, saio98, saio04, kondo}.  When dynamical carbon burning does not
occur in the merger phase and the total mass of the system does not exceed
$M_{\rm Ch}$, the system would form a massive CO WD. We call this evolutionary
path as the massive white dwarf (MWD) path.

\begin{figure}
  \begin{center}

    \begin{minipage}{0.95\hsize}
      \begin{center}
        \includegraphics[width=7.5cm, angle=0]{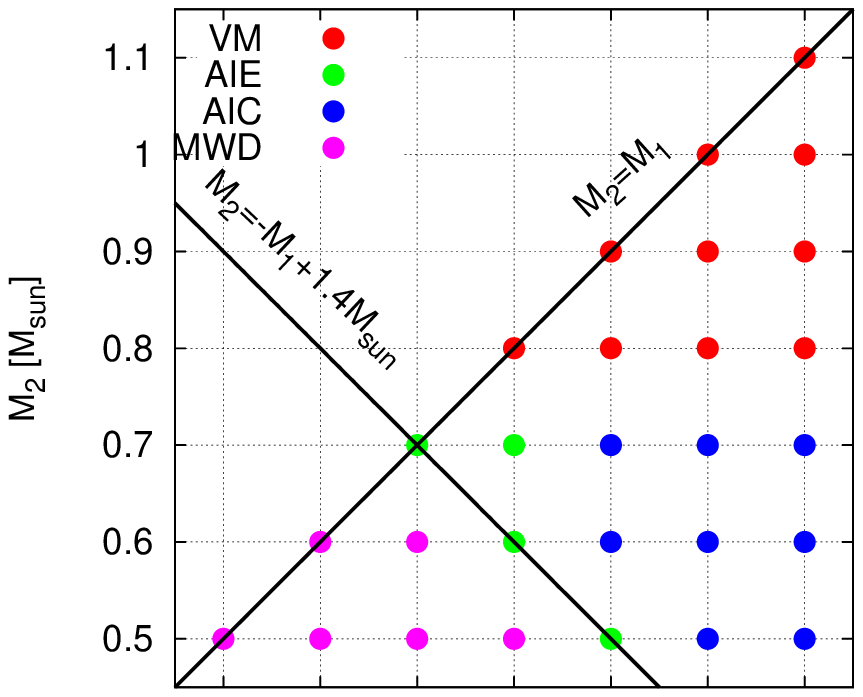}
      \end{center}
    \end{minipage}

    \begin{minipage}{0.95\hsize}
      \begin{center}
        \includegraphics[width=7.5cm, angle=0]{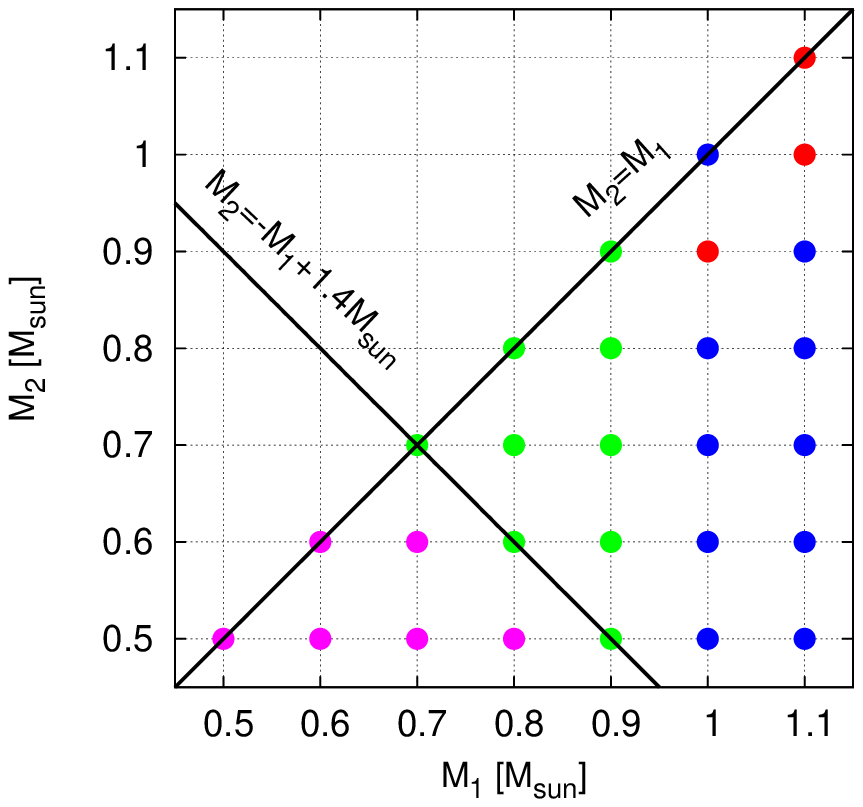}
      \end{center}
    \end{minipage}

  \end{center}
  \vspace{10pt}
 \caption{The outcome of our merger simulations for the (a) raw
and (b) smoothed temperatures.  Red symbols mean 
the VM path, green symbols the AIE path, blue the AIC, and magenta
the MWD path.  Two black lines indicate that the primary and secondary
have the same mass and that the total mass equals $M_1+M_2=M_{\rm Ch}
\sim 1.4 ~M_{\odot}$.  Mass combinations of red and
green symbols result in an SN Ia.}
 \label{fig.4-3}
\end{figure}

Figure \ref{fig.4-3} shows all mass combinations of our simulations with
$500k~M_{\odot}^{-1}$ particles and identifies which path they take.
Colors of symbols indicate the four paths, i.e., the VM (red), AIE (green),
AIC (blue), and MWD (magenta) paths.  Among these four paths,
the VM and AIE paths are possible paths to SNe Ia.

\subsection{SN Ia Rate}
\label{sn1a_rate}

We estimate the rate of SNe Ia which originate from CO WD mergers and
their contribution to the entire SNe Ia in our galaxy. Here, we assume
that all mergers of CO WDs satisfying the VM or AIE condition
can explode as an SN Ia. Strictly speaking,
for the VM path, dynamical carbon burning is a necessary,
but not a sufficient condition for carbon detonation.
So it is not trivial if those mergers explode as an SN Ia. 
On the other hand, we have to follow the evolution of merger remnants
for a much longer time \citep[e.g.,][]{yoon} in order to identify
their final fates, the AIE (exploding as an SN Ia) or 
AIC (collapsing to a neutron star) path.  
In this sense, our estimate is just an upper limit.

The SN Ia rate depends on the Hubble type and stellar mass of a galaxy.
According to \citet{li}, in SBc type galaxies with similar stellar 
mass to our galaxy, the SN Ia rate is about 
$1.1{\times}10^{-13}$~yr$^{-1}~M_{\odot}^{-1}$.
\citet{badenes} estimated the merger rate of binary WDs in our galaxy as
$1.4{\times}10^{-13}$~yr$^{-1}~M_{\odot}^{-1}$. 
The mass distribution of binary WDs in our galaxy is still uncertain
because there are small samples even in our neighborhood. We assume
that both the primary and the secondary follow the mass distribution
of single DA WDs in our galaxy derived from SDSS-DR7 
\citep[see, e.g., Figure 10 of][]{kleinman}.
Then we can calculate the rate of CO WD mergers that
satisfy the condition of each scenario.

The merger rate is about $1.4{\times}10^{-15}$~yr$^{-1} M_{\odot}^{-1}$
($0.14{\times}10^{-15}$~yr$^{-1} M_{\odot}^{-1}$) for the VM, 
$5.4{\times}10^{-15}$~yr$^{-1} M_{\odot}^{-1}$ 
($8.9{\times}10^{-15}$~yr$^{-1} M_{\odot}^{-1}$) for the AIE path,
and $6.8{\times}10^{-15}$~yr$^{-1} M_{\odot}^{-1}$ 
($9.0{\times}10^{-15}$~yr$^{-1} M_{\odot}^{-1}$) for the both paths,
if we adopt the case of raw temperature (smoothed temperature).
This is only ${\ltsim} 9\%$ of the entire galactic SNe Ia. Therefore,
at least in our galaxy, DD merger systems might not dominate
the progenitors of SNe Ia.

\section{Discussion}
\label{discussion}

\subsection{Comparison With Previous Studies}
\label{comparison}

We compare the present results with previous studies. In particular,
we mainly focus on two mass combinations.  One is
$1.1+0.9 ~M_{\odot}$ and the other is $0.9+0.6 ~M_{\odot}$,
because these two combinations were well studied in the previous works.

\subsubsection{$1.1+0.9 ~M_{\odot}$}

\begin{figure}
 \begin{center}
  \includegraphics[width=7.5cm, angle=0]{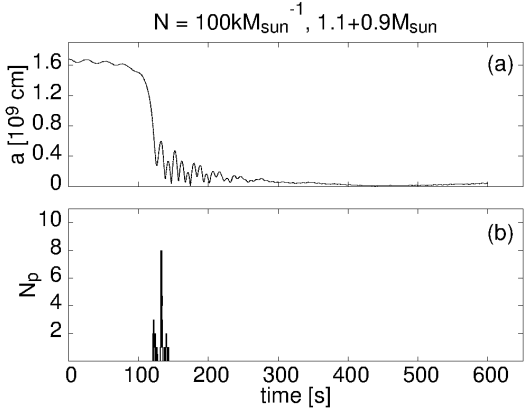}
 \end{center}
 \vspace{5pt}
 \caption{Time evolution of (a) the orbital separation and (b) the number
of high temperature (${\geq} 2.0{\times}10^{9}${\,}K) particles. The mass combination of WDs 
is $1.1+0.9 ~M_{\odot}$ and the numerical resolution is $100k~M_{\odot}^{-1}$.}
 \label{fig.4-1}
\end{figure}

First we compare our result of $1.1+0.9 ~M_{\odot}$ and $100k~M_{\odot}^{-1}$
(i.e. the total number of SPH particles is about $2{\times}10^5$)
with that of \citet{pakmor12b}.  
Figure \ref{fig.4-1}(a) shows the time evolutions of the orbital separation
and Figure \ref{fig.4-1}(b) shows the number of particles having
temperature higher than $2{\times}10^{9}~$K.
Our Figure \ref{fig.4-1} should be compared with
Figure 4 and Table 1 of \citet{pakmor12b}.

It should be noted that the WDs in our model merge more quickly (about 
100~s) than those in \citet{pakmor12b} (about 600~s) although we start
our simulation with the initial condition similar to theirs.
This is because our initial separation (${\sim} 1.67{\times}10^9~{\rm cm}$) 
is less than \citet{pakmor12b} (${\sim} 1.93{\times}10^9~{\rm cm}$).
Indeed, Figure \ref{fig.4-1}(a) and \ref{fig.4-1}(b) resemble the case of 
their smaller initial separation.
We suppose that the main reason of the difference
is the relaxation method of a single WD.
As mentioned in Section \ref{initial_setup}, we relax a single WD with 
a velocity-dependent damping force but without the evolution of
internal energy.  On the other hand, \citet{pakmor12b}
changed the damping timescale (see their Equation 14).
After the relaxation method of \citet{pakmor12b} is applied to our models, 
the radii of the relaxed WDs are a few percent larger than our original ones.
Therefore, their separations when the RLOF starts are larger than ours
and it results in a longer merging time.

If the initial separation is smaller when the RLOF starts, 
the mass transfer tends to occur more violently 
and the secondary is completely disrupted in 
a few orbital periods \citep{dan11}.
As a result, accreted matter is more strongly
shock-heated and dynamical carbon burning easily occurs.
Although \citet{pakmor12b} concluded that the initial condition
is not so important for dynamical carbon burning, we should note
that our highest temperatures in the merger phase 
could be overestimated.

\subsubsection{$0.9+0.6{M_{\odot}}$}
\label{09_06m}

\begin{figure}
 \begin{center}
  \includegraphics[width=7.5cm, angle=0]{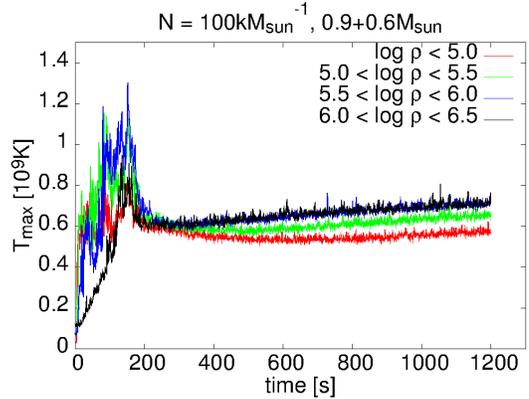}
 \end{center}
 \vspace{5pt}
 \caption{Evolution of highest temperature in each density range,
$\log {\rho}~({\rm g~cm}^{-3}) < 5.0$ (red), 
$5.0 < \log {\rho} < 5.5$ (green), $5.5 < \log 
{\rho} < 6.0$ (blue), $6.0 < \log {\rho} < 6.5$ (black), for the mass
combination of $0.9+0.6 ~M_{\odot}$ and the resolution of $100k~M_{\odot}^{-1}$.}
 \label{fig.4-2}
\end{figure}

We also compare our result of $0.9+0.6 ~M_{\odot}$ and $100k~M_{\odot}^{-1}$
with the previous studies of \citet{yoon}, \citet{dan11}, and \citet{zhu}.
Since these studies have similar resolution (a few ${\times}10^5$ 
SPH particles in all), it is suitable for comparison.
Figure \ref{fig.4-2} shows the time evolution of highest temperature
in each density zone.
Since no dynamical carbon burning occurs (see Figure \ref{fig.4-3}), 
we examine whether or not quiescent carbon burning occurs
in the remnant phase, i.e., at $t \sim300${\,}s in Figure \ref{fig.4-2}.
The temperature is about $6{\times}10^{8}${\,}K and this is consistent with
the above three previous studies. 

We compare our Figure \ref{fig.4-2} with Figure 4 in \citet{yoon}.
For the highest temperature in the merger phase (at $t\sim100${\,}s
in Figure \ref{fig.4-2}), their results of $1.7{\times}10^{9}${\,}K is
higher than ours ($1.3{\times}10^{9}${\,}K), although dynamical carbon
burning does not occur in the merger phase for both ours and theirs.
We suppose that the difference in the highest temperature comes from
the difference in the initial condition.  
\citet{dan11} reported that the morphology of merger remnant could
be affected by the initial condition.
They found that the remnant whose initial separation is larger has
longer trailing arm than the one whose initial separation is smaller
\citep[see Figure 10 of][]{dan11}.  On the other hand, we find
that the highest temperature in the remnant phase
is barely affected by the initial condition (see Figure \ref{fig.4-2}).
Similar discussion appears in \citet{tanikawa}.

Other factors might affect the results of our simulations. 
For example, \citet{zhu} and \citet{dan14} performed merger simulations 
of non-spinning WDs and found that the structures of such merger 
remnants are different from those in mergers of synchronously spinning WDs,
especially for the case of nearly equal masses.  
In non-spinning cases, the high temperature region 
is formed near the center of the merger remnant, and 
carbon burning might occur in that region.
Since it is still uncertain whether binary WDs
maintain synchronization until their merger, our results for the AIE path
might be changed.  The topic of synchronization is out of the scope
in this paper, so we leave this in our future works.

\subsection{Numerical Resolution}
\label{numerical_resolution}

\citet{pakmor12b} concluded that the numerical resolution of simulation is
one of the most important factors for carbon burning in the merger phase
because very high resolution is required to identify very small hot spots.
Therefore, we examine the dependence of our results on the numerical
resolution.
We perform the same simulations with four different resolutions, i.e.,
$10k$, ~$50k$, ~$100k$, and ~$500k$ per one solar mass.
The highest temperature is 
critical both for dynamical carbon burning and quiescent carbon burning,
so we focus on the dependence of the highest temperature on the resolution.

\begin{figure}
 \begin{center}

    \begin{minipage}{0.9\hsize}
      \begin{center}
        \includegraphics[width=7.5cm, angle=0]{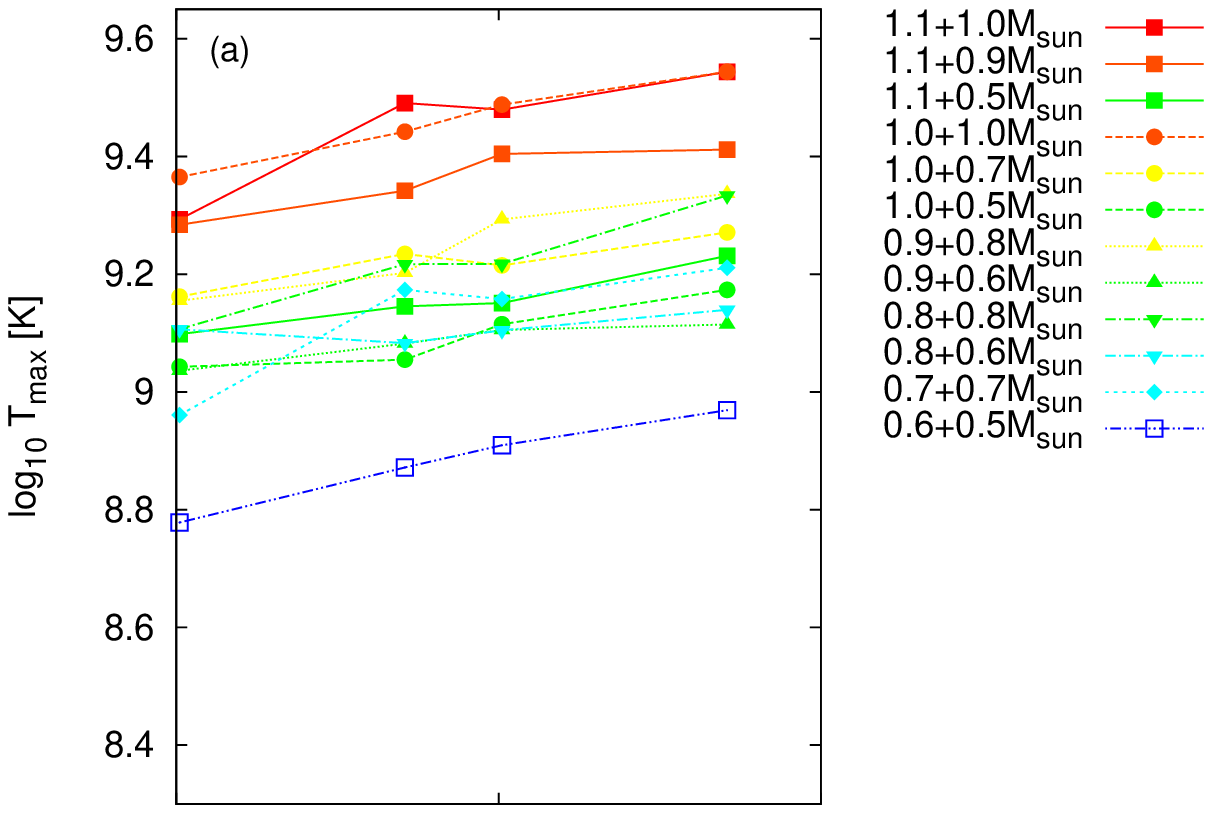}
      \end{center}
    \end{minipage}

    \begin{minipage}{0.9\hsize}
      \begin{center}
        \includegraphics[width=7.5cm, angle=0]{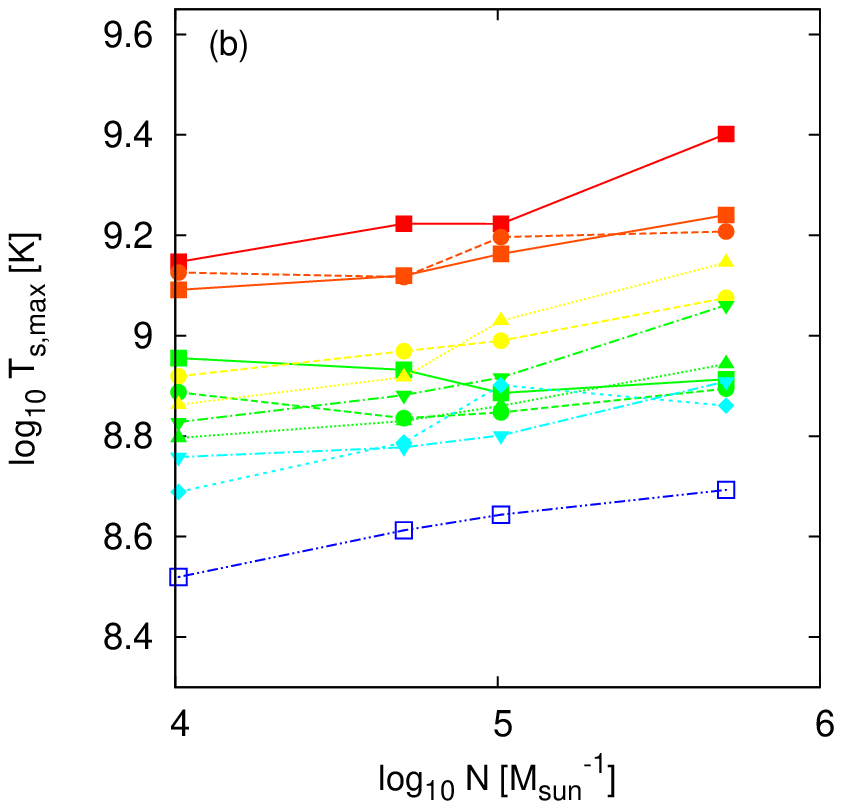}
      \end{center}
    \end{minipage}

 \end{center}
 \vspace{10pt}
 \caption{Dependence of highest temperature on the numerical resolution
in the merger phase for the (a) raw and (b) smoothed temperatures.
The horizontal axis is the number of SPH particles
per a solar mass.  The vertical axis is the highest temperature. 
Shapes and colors of symbols have the same meaning as those 
in Figure \ref{fig.3-2}.  The highest temperature tends to increase
with the numerical resolution.}
 \label{fig.4-4}
\end{figure}

\begin{figure}
 \begin{center}
   
    \begin{minipage}{0.9\hsize}
      \begin{center}
        \includegraphics[width=7.5cm, angle=0]{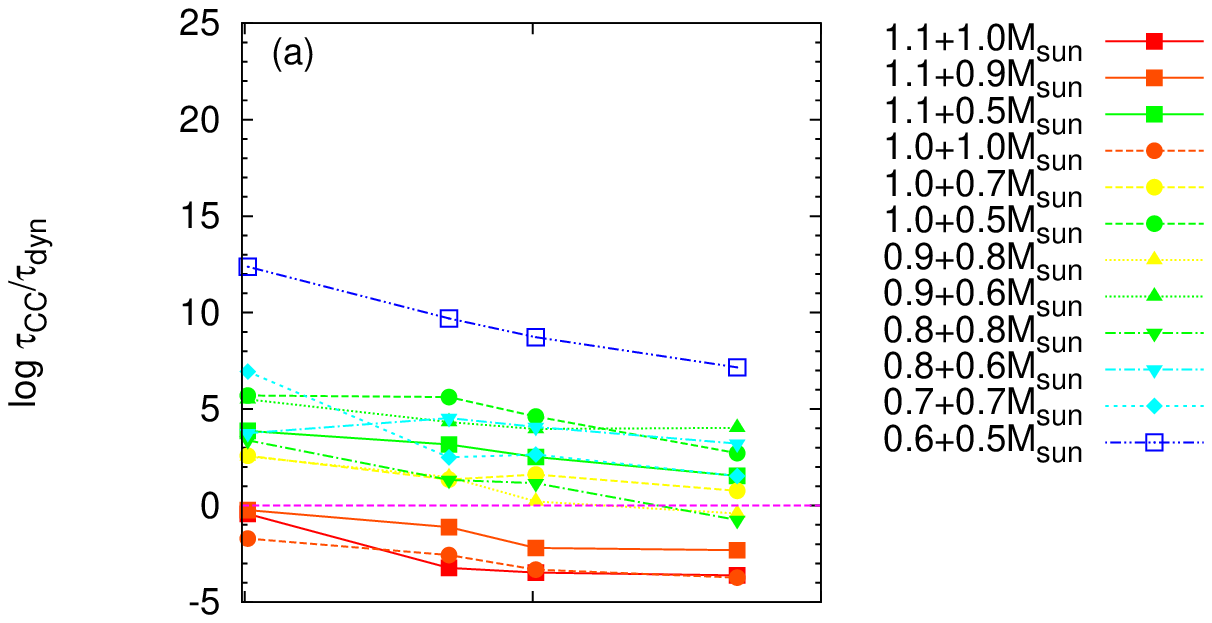}
      \end{center}
    \end{minipage}

    \begin{minipage}{0.9\hsize}
      \begin{center}
        \includegraphics[width=7.5cm, angle=0]{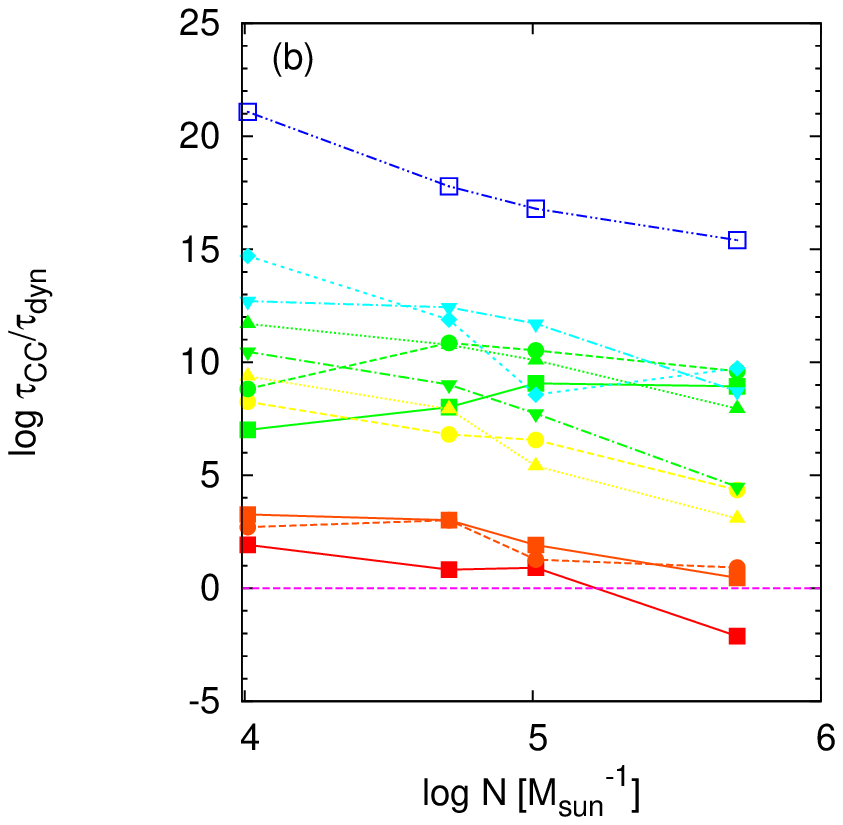}
      \end{center}
    \end{minipage}

 \end{center}
 \vspace{10pt}
 \caption{Dependence of minimum ${\tau}_{\rm CC}/{\tau}_{\rm dyn}$
ratio on the numerical resolution in the merger phase for
the (a) raw and (b) smoothed temperatures.
The horizontal magenta dashed lines indicate
${\tau}_{\rm CC}/{\tau}_{\rm dyn} = 1.0$.}
 \label{fig.4-5}
\end{figure}

Figure \ref{fig.4-4}(a) and \ref{fig.4-4}(b) show the dependence
of the highest temperature in the merger phase on the numerical resolution
for the (a) raw and (b) smoothed temperatures, respectively.
The highest temperature increases with the number of SPH particles.
In other words, our simulation does not converge yet and 
the final fates of some models could be changed from the AIC path
to the VM path.  Such a tendency was also reported in \citet{pakmor12b}.
In this sense, we must further increase the numerical resolution 
to definitely identify the fate of merger products, at least, more
than $500k~M_{\odot}^{-1}$. 
Figure \ref{fig.4-5}(a) and \ref{fig.4-5}(b) show the dependence
of the minimum ${\tau}_{\rm CC}/{\tau}_{\rm dyn}$ ratio in the merger phase 
on the numerical resolution for the (a) raw and (b) smoothed temperatures, 
respectively. 
The magenta dashed line indicates ${\tau}_{\rm CC}/{\tau}_{\rm dyn} = 1.0$. 
It tends to decrease as the numerical resolution increases.
This trend is consistent with that of the highest temperature.
For smoothed temperature, only very massive pairs 
(both masses ${\gtsim} 1.0~M_\sun$) can ignite carbon dynamically.

\begin{figure}
 \begin{center}
   
    \begin{minipage}{0.9\hsize}
      \begin{center}
        \includegraphics[width=7.5cm, angle=0]{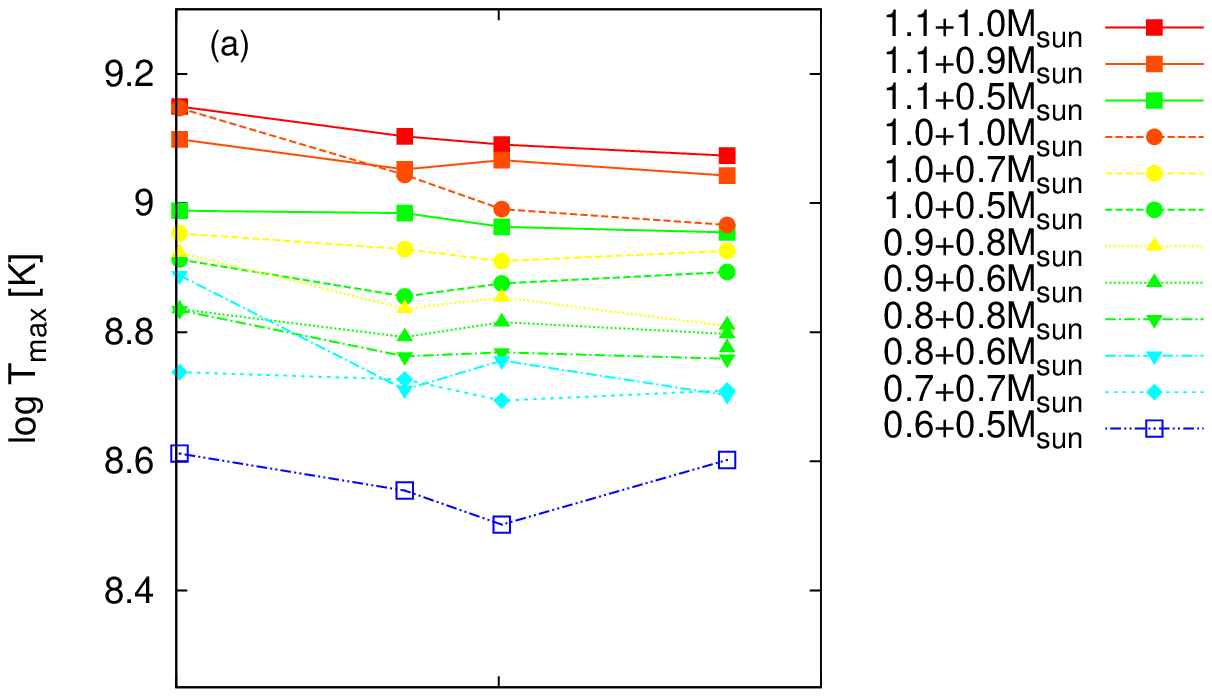}
      \end{center}
    \end{minipage}

    \begin{minipage}{0.9\hsize}
      \begin{center}
        \includegraphics[width=7.5cm, angle=0]{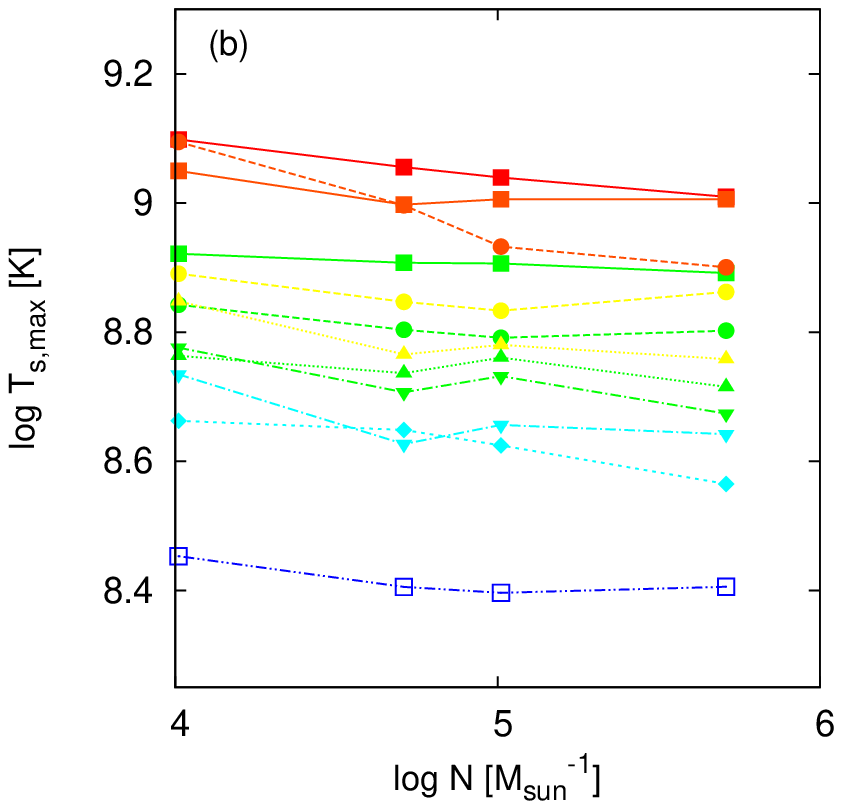}
      \end{center}
    \end{minipage}

 \end{center}
 \vspace{10pt}
 \caption{Same as Figure \ref{fig.4-4}, but for the highest temperature
in the remnant phase.  The highest temperature seems to converge.}
 \label{fig.4-6}
\end{figure}

On the other hand, Figure \ref{fig.4-6}(a) and \ref{fig.4-6}(b)
show the highest temperature in the remnant phase
for the (a) raw and (b) smoothed temperatures, respectively.
Comparing with the results in the merger phase (Figure \ref{fig.4-4}),
the highest temperature depends barely on the resolution.  Especially,
for the smoothed temperature in Figure \ref{fig.4-6}(b), it 
converges for almost all the mass combinations.
This tendency of weak dependence on the resolution was already 
reported in the previous studies \citep{raskin12, dan14}.

\section{Conclusions}
\label{conclusions}

We have performed SPH simulations of CO WD mergers
for the mass combinations of
$0.5$--$1.1 ~M_{\odot}$ from the start of the RLOF to 
the formation of a quasi-stationary merger remnant,
and examined whether carbon burning occurs either
in the merger phase or remnant phase.  Using the results of SPH simulations,
we have investigated the mass range of CO WDs that 
possibly lead to an SN Ia in the merger phase or remnant phase.
We have obtained the mass range as follows.
When the primary and secondary are
as massive as $0.9~M_{\odot} {\le} M_{1,2} {\le} 1.1 ~M_{\odot}$,
the binary results in an SN Ia in the merger phase.
On the other hand, when the primary is $0.7~M_\odot \le M_1 \le 0.9 ~M_\odot$
and the total mass of the binary exceeds $1.4 ~M_\odot$,
they lead to an SN Ia in the remnant phase.

From the obtained mass range, 
we have estimated the rate of SNe Ia 
coming from CO WD mergers in our galaxy.
It is $6.8{\times}10^{-15}$~yr$^{-1}~M_{\odot}^{-1}$ if we use the results
of our raw temperature calculations, while it is 
$9.0{\times}10^{-15}$~yr$^{-1} ~M_{\odot}^{-1}$ if we use that of
the smoothed temperature.
These are only less than 9\% of
the entire SN Ia rate. 
Therefore, it is unlikely that the mergers 
of CO WDs are the main progenitors of SNe Ia.

Of course, the above estimate is not conclusive
because of several uncertainties in our calculation.
We have checked the dependence of the highest temperature on the numerical resolution
both in the merger and remnant phases in order to examine the numerical
convergence of our simulations.  We have found that the highest temperature
in the merger phase depends on the numerical resolution.  It tends
to increase with the resolution as already reported in \citet{pakmor12b}. 
On the other hand, in the remnant phase, 
the highest temperature depends barely on the numerical resolution.
Therefore, it is necessary to increase the number of SPH particles,
at least, up to ${\ge}500k~M_\odot^{-1}$, 
for definite conclusion.
Additionally, our calculations for SNe Ia 
in the remnant phase is not sufficient.
In order to obtain the decisive conclusion,
we have to follow the further evolution of the merger remnant,
like \citet{yoon}.
This is one of our future works.\\


We thank the anonymous referee for many detailed comments that help
to improve the paper.
These simulations were performed by using computational resources of
Kavli Institute for the Physics and Mathematics of the Universe (IPMU),
and HA-PACS at the Center for Computational Sciences in University of
Tsukuba under Interdisciplinary Computational Science Program.
This research has been supported in part by Grants-in-Aid for Scientific
Research (23224004, 23540262, 23740141, 24540227, 26400222, and 26800100) 
from the Japan Society for the Promotion of Science and by the World Premier
International Research Center Initiative, MEXT, Japan.
This work is partly supported by MEXT program for the Development and Improvement 
for the Next Generation Ultra High-Speed Computer System under its Subsidies
for Operating the Specific Advanced Large Research Facilities.

\appendix

\section{Summary of CO WD merger simulations}
\label{summary_appendix}

We summarize our numerical results in Table 1 and 2.
The numeric values presented in the table are the primary mass,
$M_1$, secondary mass, $M_2$, initial orbital separation, $a_{\rm init}$,
raw and smoothed maximum temperatures, $T_{\rm max}$ and $T_{{\rm s,max}}$,
densities at the highest raw temperature and at the highest smoothed
temperature, $\rho(T_{\rm max})$ and $\rho(T_{{\rm s,max}})$, in the 
merger phase, $T_{\rm max,rem}$ and $T_{{\rm s,max,rem}}$, 
$\rho(T_{\rm max,rem})$ and $\rho(T_{{\rm s,max,rem}})$ in the
remnant phase.


\newpage
\begin{deluxetable}{ccccccccccc}
\tabletypesize{\scriptsize}
\tablecaption{Summary of all calculated models
\label{tab.4-1}}
\tablewidth{0pt}
\tablehead{
\colhead{$M_1$} & \colhead{$M_2$} & \colhead{$a_{\rm init}$}
& \colhead{$T_{\rm max}$} & \colhead{$\rho(T_{\rm max})$}
& \colhead{$T_{\rm s,max}$} & \colhead{$\rho(T_{\rm s, max})$}
& \colhead{$T_{\rm max,rem}$} & \colhead{$\rho(T_{\rm max,rem})$}
& \colhead{$T_{\rm s,max,rem}$} 
& \colhead{$\rho(T_{\rm s,max,rem})$} \\
\colhead{$(M_\odot)$} & \colhead{$(M_\odot)$} & \colhead{($10^9$ cm)}
& \colhead{($10^9$ K)} & \colhead{($10^6$ g cm$^{-3}$)}
& \colhead{($10^9$ K)} & \colhead{($10^6$ g cm$^{-3}$)} 
& \colhead{($10^9$ K)} & \colhead{($10^6$ g cm$^{-3}$)} 
& \colhead{($10^9$ K)} & \colhead{($10^6$ g cm$^{-3}$)}
}
\startdata
\hline
      \multicolumn{11}{c} {$\mathrm{Resolution} = 10k~\mathrm{M_{\odot}}^{-1}$} \\ \hline
      1.1 & 1.1 & 1.22 & 2.66 & 6.72 & 1.94 & 7.57 & 1.81 & 2.72 & 1.69 & 3.12\\
      1.1 & 1.0 & 1.42 & 1.96 & 3.65 & 1.40 & 9.52 & 1.41 & 1.94 & 1.25 & 2.52\\
      1.1 & 0.9 & 1.64 & 1.92 & 3.37 & 1.23 & 6.59 & 1.25 & 3.02 & 1.12 & 2.98\\
      1.1 & 0.8 & 1.86 & 1.86 & 1.83 & 1.10 & 4.00 & 1.18 & 2.94 & 1.02 & 2.37\\
      1.1 & 0.7 & 2.11 & 1.45 & 0.87 & 1.02 & 3.90 & 1.25 & 1.82 & 0.94 & 2.74\\
      1.1 & 0.6 & 2.40 & 1.40 & 1.21 & 0.92 & 3.20 & 1.01 & 2.64 & 0.88 & 2.48\\
      1.1 & 0.5 & 2.75 & 1.25 & 1.10 & 0.90 & 1.68 & 0.97 & 1.03 & 0.83 & 2.29\\
      1.0 & 1.0 & 1.40 & 2.31 & 4.67 & 1.33 & 4.11 & 1.40 & 3.12 & 1.24 & 3.03\\
      1.0 & 0.9 & 1.61 & 1.54 & 3.24 & 0.94 & 3.26 & 1.05 & 2.35 & 0.94 & 2.16\\
      1.0 & 0.8 & 1.81 & 1.73 & 2.16 & 0.88 & 2.93 & 0.93 & 2.50 & 0.83 & 2.01\\
      1.0 & 0.7 & 2.05 & 1.45 & 1.08 & 0.83 & 0.75 & 0.90 & 2.12 & 0.78 & 2.51\\
      1.0 & 0.6 & 2.35 & 1.35 & 1.10 & 0.79 & 1.01 & 0.82 & 2.78 & 0.74 & 2.52\\
      1.0 & 0.5 & 2.68 & 1.10 & 0.32 & 0.77 & 1.50 & 0.82 & 3.08 & 0.70 & 1.88\\
      0.9 & 0.9 & 1.62 & 1.75 & 2.12 & 0.89 & 1.20 & 0.99 & 2.83 & 0.89 & 1.20\\
      0.9 & 0.8 & 1.79 & 1.43 & 1.52 & 0.73 & 2.02 & 0.84 & 2.58 & 0.70 & 1.55\\
      0.9 & 0.7 & 2.06 & 1.23 & 1.91 & 0.73 & 1.30 & 0.72 & 2.07 & 0.61 & 1.98\\
      0.9 & 0.6 & 2.34 & 1.09 & 0.62 & 0.63 & 0.65 & 0.68 & 1.17 & 0.58 & 1.37\\
      0.9 & 0.5 & 2.67 & 1.03 & 0.82 & 0.59 & 0.42 & 0.65 & 2.48 & 0.53 & 1.15\\
      0.8 & 0.8 & 1.77 & 1.28 & 2.19 & 0.67 & 1.53 & 0.68 & 1.89 & 0.60 & 1.30\\
      0.8 & 0.7 & 2.00 & 1.27 & 1.75 & 0.61 & 1.49 & 0.57 & 0.87 & 0.49 & 1.41\\
      0.8 & 0.6 & 2.26 & 1.28 & 0.99 & 0.57 & 0.93 & 0.77 & 1.20 & 0.54 & 1.20\\
      0.8 & 0.5 & 2.58 & 0.84 & 0.42 & 0.53 & 1.27 & 0.57 & 0.93 & 0.45 & 0.45\\
      0.7 & 0.7 & 1.95 & 0.91 & 1.40 & 0.49 & 1.48 & 0.55 & 0.98 & 0.46 & 1.26\\
      0.7 & 0.6 & 2.21 & 1.01 & 1.42 & 0.48 & 0.96 & 0.50 & 1.08 & 0.41 & 1.08\\
      0.7 & 0.5 & 2.49 & 0.82 & 0.37 & 0.49 & 0.36 & 0.45 & 0.38 & 0.37 & 0.86\\
      0.6 & 0.6 & 2.16 & 0.72 & 1.14 & 0.36 & 0.98 & 0.45 & 0.41 & 0.32 & 0.41\\
      0.6 & 0.5 & 2.40 & 0.60 & 0.45 & 0.33 & 0.44 & 0.41 & 1.28 & 0.28 & 1.06\\
      0.5 & 0.5 & 2.35 & 0.57 & 1.06 & 0.29 & 0.65 & 0.30 & 0.76 & 0.24 & 0.49\\ \hline
      \multicolumn{11}{c} {$\mathrm{Resolution} = 50k~\mathrm{M_{\odot}}^{-1}$} \\ \hline
      1.1 & 1.1 & 1.23 & 3.58 & 6.13 & 1.81 & 8.88 & 1.56 & 3.02 & 1.44 & 3.09\\
      1.1 & 1.0 & 1.44 & 3.10 & 4.15 & 1.67 & 3.59 & 1.27 & 2.71 & 1.14 & 2.83\\
      1.1 & 0.9 & 1.65 & 2.20 & 2.87 & 1.32 & 2.66 & 1.13 & 2.94 & 0.99 & 2.36\\
      1.1 & 0.8 & 1.89 & 1.94 & 1.47 & 1.16 & 1.10 & 1.08 & 2.42 & 0.93 & 2.80\\
      1.1 & 0.7 & 2.13 & 1.58 & 0.39 & 1.06 & 3.90 & 1.00 & 1.96 & 0.86 & 1.76\\
      1.1 & 0.6 & 2.42 & 1.52 & 0.33 & 0.88 & 0.51 & 0.94 & 1.89 & 0.82 & 2.19\\
      1.1 & 0.5 & 2.79 & 1.40 & 0.67 & 0.86 & 0.60 & 0.96 & 1.40 & 0.81 & 1.49\\
      1.0 & 1.0 & 1.42 & 2.77 & 3.55 & 1.31 & 3.10 & 1.11 & 2.98 & 0.99 & 3.08\\
      1.0 & 0.9 & 1.62 & 2.16 & 2.67 & 1.06 & 3.01 & 0.94 & 1.87 & 0.81 & 2.73\\
      1.0 & 0.8 & 1.85 & 1.93 & 1.98 & 1.05 & 1.10 & 0.86 & 1.57 & 0.76 & 1.29\\
      1.0 & 0.7 & 2.09 & 1.72 & 0.93 & 0.93 & 1.16 & 0.85 & 0.86 & 0.70 & 1.31\\
      1.0 & 0.6 & 2.37 & 1.28 & 0.50 & 0.78 & 0.56 & 0.79 & 0.88 & 0.66 & 2.01\\
      1.0 & 0.5 & 2.73 & 1.13 & 0.23 & 0.69 & 0.31 & 0.72 & 1.14 & 0.64 & 1.41\\
      0.9 & 0.9 & 1.62 & 2.15 & 2.22 & 1.07 & 2.17 & 0.77 & 2.35 & 0.66 & 1.55\\
      0.9 & 0.8 & 1.84 & 1.59 & 2.08 & 0.83 & 1.92 & 0.69 & 1.55 & 0.58 & 0.85\\
      0.9 & 0.7 & 2.09 & 1.44 & 0.79 & 0.81 & 1.27 & 0.64 & 0.44 & 0.55 & 0.44\\
      0.9 & 0.6 & 2.37 & 1.21 & 0.86 & 0.68 & 0.58 & 0.62 & 1.29 & 0.55 & 0.70\\
      0.9 & 0.5 & 2.73 & 1.00 & 0.29 & 0.59 & 0.18 & 0.63 & 2.51 & 0.50 & 0.37\\
      0.8 & 0.8 & 1.80 & 1.65 & 1.62 & 0.76 & 1.37 & 0.58 & 1.55 & 0.51 & 1.91\\
      0.8 & 0.7 & 2.03 & 1.59 & 1.39 & 0.76 & 1.31 & 0.56 & 2.81 & 0.45 & 0.68\\
      0.8 & 0.6 & 2.30 & 1.21 & 0.56 & 0.60 & 0.40 & 0.51 & 0.62 & 0.42 & 0.62\\
      0.8 & 0.5 & 2.65 & 0.96 & 0.29 & 0.52 & 0.30 & 0.49 & 0.73 & 0.41 & 0.38\\
      0.7 & 0.7 & 2.00 & 1.50 & 0.84 & 0.61 & 0.80 & 0.53 & 1.49 & 0.45 & 1.07\\
      0.7 & 0.6 & 2.29 & 1.10 & 1.04 & 0.53 & 0.80 & 0.45 & 2.94 & 0.34 & 0.32\\
      0.7 & 0.5 & 2.61 & 0.98 & 0.33 & 0.45 & 0.11 & 0.42 & 0.32 & 0.33 & 0.32\\
      0.6 & 0.6 & 2.22 & 0.93 & 0.81 & 0.40 & 0.86 & 0.48 & 2.78 & 0.31 & 0.55\\
      0.6 & 0.5 & 2.52 & 0.74 & 0.48 & 0.41 & 0.36 & 0.36 & 0.32 & 0.25 & 0.32\\
      0.5 & 0.5 & 2.44 & 0.85 & 0.59 & 0.37 & 0.46 & 0.35 & 1.37 & 0.23 & 0.27\\

\enddata
\end{deluxetable}

\begin{deluxetable}{ccccccccccc}
\tabletypesize{\scriptsize}
\tablecaption{Summary of all calculated models
\label{tab.4-2}}
\tablewidth{0pt}
\tablehead{
\colhead{$M_1$} & \colhead{$M_2$} & \colhead{$a_{\rm init}$}
& \colhead{$T_{\rm max}$} & \colhead{$\rho(T_{\rm max})$}
& \colhead{$T_{\rm s,max}$} & \colhead{$\rho(T_{\rm s, max})$}
& \colhead{$T_{\rm max,rem}$} & \colhead{$\rho(T_{\rm max,rem})$}
& \colhead{$T_{\rm s,max,rem}$} 
& \colhead{$\rho(T_{\rm s,max,rem})$} \\
\colhead{$(M_\odot)$} & \colhead{$(M_\odot)$} & \colhead{($10^9$ cm)}
& \colhead{($10^9$ K)} & \colhead{($10^6$ g cm$^{-3}$)}
& \colhead{($10^9$ K)} & \colhead{($10^6$ g cm$^{-3}$)} 
& \colhead{($10^9$ K)} & \colhead{($10^6$ g cm$^{-3}$)} 
& \colhead{($10^9$ K)} & \colhead{($10^6$ g cm$^{-3}$)}
}
\startdata
\hline
      \multicolumn{11}{c} {$\mathrm{Resolution} = 100k~\mathrm{M_{\odot}}^{-1}$} \\ \hline

      1.1 & 1.1 & 1.25 & 3.64 & 5.51 & 1.79 & 3.38 & 1.39 & 2.82 & 1.25 & 2.88\\
      1.1 & 1.0 & 1.45 & 3.02 & 6.71 & 1.67 & 3.05 & 1.23 & 2.67 & 1.10 & 3.11\\
      1.1 & 0.9 & 1.67 & 2.54 & 4.13 & 1.46 & 4.31 & 1.17 & 2.74 & 1.01 & 1.73\\
      1.1 & 0.8 & 1.86 & 2.21 & 1.92 & 1.28 & 1.67 & 1.08 & 2.12 & 0.96 & 2.96\\
      1.1 & 0.7 & 2.11 & 1.82 & 1.10 & 1.09 & 0.90 & 0.98 & 2.47 & 0.90 & 1.51\\
      1.1 & 0.6 & 2.39 & 1.56 & 0.42 & 0.91 & 0.65 & 0.95 & 1.82 & 0.82 & 1.24\\
      1.1 & 0.5 & 2.72 & 1.42 & 1.91 & 0.77 & 0.88 & 0.92 & 0.67 & 0.81 & 0.83\\
      1.0 & 1.0 & 1.43 & 3.08 & 4.80 & 1.57 & 4.17 & 0.98 & 3.05 & 0.86 & 2.57\\
      1.0 & 0.9 & 1.64 & 2.10 & 3.32 & 1.17 & 2.14 & 0.93 & 1.15 & 0.80 & 1.70\\
      1.0 & 0.8 & 1.87 & 1.99 & 2.49 & 1.17 & 1.07 & 0.84 & 0.73 & 0.74 & 0.97\\
      1.0 & 0.7 & 2.13 & 1.64 & 1.05 & 0.98 & 0.62 & 0.81 & 0.38 & 0.68 & 0.45\\
      1.0 & 0.6 & 2.40 & 1.45 & 1.23 & 0.89 & 0.76 & 0.77 & 0.98 & 0.66 & 1.16\\
      1.0 & 0.5 & 2.75 & 1.30 & 0.18 & 0.70 & 0.32 & 0.75 & 0.90 & 0.62 & 0.78\\
      0.9 & 0.9 & 1.64 & 2.17 & 3.21 & 0.95 & 2.34 & 0.78 & 2.92 & 0.66 & 2.09\\
      0.9 & 0.8 & 1.86 & 1.96 & 1.24 & 1.07 & 1.05 & 0.71 & 1.26 & 0.60 & 0.97\\
      0.9 & 0.7 & 2.08 & 1.48 & 1.51 & 0.84 & 0.56 & 0.66 & 0.40 & 0.59 & 0.38\\
      0.9 & 0.6 & 2.35 & 1.27 & 0.67 & 0.73 & 0.39 & 0.65 & 0.56 & 0.58 & 0.72\\
      0.9 & 0.5 & 2.68 & 1.07 & 0.31 & 0.61 & 0.18 & 0.59 & 0.48 & 0.50 & 0.50\\
      0.8 & 0.8 & 1.83 & 1.65 & 2.20 & 0.83 & 3.35 & 0.59 & 2.92 & 0.54 & 1.31\\
      0.8 & 0.7 & 2.07 & 1.51 & 0.89 & 0.81 & 0.82 & 0.58 & 0.79 & 0.46 & 0.97\\
      0.8 & 0.6 & 2.35 & 1.27 & 0.56 & 0.63 & 0.43 & 0.57 & 0.43 & 0.45 & 0.37\\
      0.8 & 0.5 & 2.69 & 1.19 & 0.37 & 0.59 & 0.19 & 0.50 & 2.74 & 0.40 & 0.51\\
      0.7 & 0.7 & 2.01 & 1.44 & 1.12 & 0.80 & 1.10 & 0.49 & 0.86 & 0.42 & 0.79\\
      0.7 & 0.6 & 2.27 & 1.12 & 0.45 & 0.63 & 0.39 & 0.47 & 0.60 & 0.35 & 0.46\\
      0.7 & 0.5 & 2.60 & 0.84 & 0.45 & 0.49 & 0.30 & 0.46 & 0.35 & 0.34 & 0.32\\
      0.6 & 0.6 & 2.25 & 1.13 & 0.76 & 0.58 & 0.67 & 0.49 & 2.61 & 0.28 & 0.43\\
      0.6 & 0.5 & 2.55 & 0.81 & 0.44 & 0.44 & 0.30 & 0.32 & 2.64 & 0.25 & 0.35\\
      0.5 & 0.5 & 2.47 & 0.85 & 0.52 & 0.42 & 0.47 & 0.40 & 1.89 & 0.22 & 0.32\\ \hline
      \multicolumn{11}{c} {$\mathrm{Resolution} = 500k~\mathrm{M_{\odot}}^{-1}$} \\ \hline
      1.1 & 1.1 & 1.25 & 3.91 & 5.84 & 2.28 & 5.27 & 1.18 & 3.10 & 0.99 & 2.90\\
      1.1 & 1.0 & 1.46 & 3.50 & 3.81 & 2.52 & 3.87 & 1.18 & 2.41 & 1.02 & 1.63\\
      1.1 & 0.9 & 1.67 & 2.58 & 4.24 & 1.74 & 3.96 & 1.10 & 1.48 & 1.01 & 1.11\\
      1.1 & 0.8 & 1.91 & 2.19 & 2.23 & 1.43 & 1.67 & 1.04 & 0.54 & 0.94 & 1.01\\
      1.1 & 0.7 & 2.16 & 1.66 & 1.50 & 1.18 & 0.69 & 0.96 & 0.99 & 0.85 & 0.91\\
      1.1 & 0.6 & 2.45 & 1.68 & 0.70 & 1.00 & 0.57 & 0.95 & 0.74 & 0.82 & 0.89\\
      1.1 & 0.5 & 2.81 & 1.70 & 0.76 & 0.82 & 0.22 & 0.90 & 0.63 & 0.78 & 0.61\\
      1.0 & 1.0 & 1.43 & 3.50 & 4.27 & 1.61 & 5.76 & 0.92 & 2.97 & 0.80 & 1.82\\
      1.0 & 0.9 & 1.64 & 2.95 & 3.78 & 1.98 & 1.81 & 0.90 & 1.89 & 0.78 & 1.77\\
      1.0 & 0.8 & 1.87 & 2.26 & 1.32 & 1.54 & 1.32 & 0.87 & 0.91 & 0.77 & 0.73\\
      1.0 & 0.7 & 2.11 & 1.87 & 0.93 & 1.19 & 1.13 & 0.84 & 1.06 & 0.73 & 0.33\\
      1.0 & 0.6 & 2.40 & 1.31 & 0.65 & 0.69 & 0.31 & 0.79 & 0.72 & 0.66 & 1.11\\
      1.0 & 0.5 & 2.74 & 1.49 & 0.58 & 0.78 & 0.15 & 0.78 & 0.74 & 0.63 & 0.51\\
      0.9 & 0.9 & 1.64 & 2.77 & 3.84 & 1.36 & 2.94 & 0.79 & 2.66 & 0.56 & 1.19\\
      0.9 & 0.8 & 1.86 & 2.17 & 1.26 & 1.40 & 0.77 & 0.64 & 0.87 & 0.57 & 1.03\\
      0.9 & 0.7 & 2.11 & 1.72 & 0.50 & 1.05 & 0.85 & 0.64 & 0.38 & 0.57 & 0.40\\
      0.9 & 0.6 & 2.40 & 1.30 & 0.43 & 0.88 & 0.36 & 0.63 & 0.51 & 0.52 & 0.46\\
      0.9 & 0.5 & 2.75 & 1.28 & 0.24 & 0.70 & 0.20 & 0.59 & 0.60 & 0.50 & 0.36\\
      0.8 & 0.8 & 1.82 & 2.16 & 2.08 & 1.15 & 1.71 & 0.57 & 1.86 & 0.47 & 0.44\\
      0.8 & 0.7 & 2.06 & 1.74 & 0.80 & 0.96 & 0.73 & 0.50 & 0.82 & 0.43 & 0.53\\
      0.8 & 0.6 & 2.34 & 1.38 & 0.75 & 0.81 & 0.46 & 0.50 & 0.54 & 0.44 & 0.24\\
      0.8 & 0.5 & 2.67 & 1.04 & 0.33 & 0.54 & 0.11 & 0.50 & 0.49 & 0.41 & 0.17\\
      0.7 & 0.7 & 2.05 & 1.63 & 1.38 & 0.73 & 0.93 & 0.51 & 1.71 & 0.37 & 0.59\\
      0.7 & 0.6 & 2.31 & 1.11 & 0.55 & 0.57 & 0.52 & 0.48 & 3.13 & 0.33 & 0.22\\
      0.7 & 0.5 & 2.65 & 1.07 & 0.25 & 0.49 & 0.19 & 0.44 & 2.57 & 0.34 & 0.15\\
      0.6 & 0.6 & 2.24 & 1.45 & 1.21 & 0.79 & 0.64 & 0.50 & 1.90 & 0.27 & 0.49\\
      0.6 & 0.5 & 2.55 & 0.93 & 0.52 & 0.49 & 0.16 & 0.40 & 1.77 & 0.25 & 0.10\\
      0.5 & 0.5 & 2.47 & 1.09 & 0.52 & 0.56 & 0.29 & 0.41 & 1.50 & 0.21 & 0.38\\ 

\enddata
\end{deluxetable}
\end{document}